\documentclass{article}

\usepackage[utf8]{inputenc}
\usepackage{authblk}
\usepackage[sort&compress]{natbib}
\usepackage{graphicx}
\usepackage{tabularx}
\usepackage{amsmath}
\usepackage{upgreek}
\usepackage{mathtools}
\usepackage{amssymb}
\usepackage{nomencl}
\usepackage{geometry}
\usepackage{epstopdf}
\usepackage{subcaption}
\usepackage{xcolor}
\usepackage{cleveref}
\usepackage[export]{adjustbox}
\usepackage{wrapfig}
\usepackage{float}
\usepackage{placeins}

\usepackage{nameref}
\usepackage[ruled,vlined]{algorithm2e}
\usepackage{soul}
\DeclareGraphicsExtensions{.png,.pdf}
\usepackage[format=plain,font=it]{caption}
\usepackage[multiple]{footmisc}
\usepackage{bigfoot}
\usepackage{multirow}
\usepackage{lineno}

\title{A stabilized finite element formulation for investigating the influence of cerebrospinal fluid on wave propagation in brain MR elastography}

\author[1,2]{Dimitrios C. Rodopoulos \thanks{Corresponding Author: Dimitrios C. Rodopoulos,
E-mail: \textcolor{blue}{rodopoulos.dimitrios@ucy.ac.cy}}}
\author[3]{Felipe Galarce}
\author[4]{Alfonso Caiazzo}
\author[5]{Ingolf Sack}
\author[1,6]{Vasileios Vavourakis \thanks{Corresponding Author: Vasileios Vavourakis, E-mail:  \textcolor{blue}{vavourakis.vasileios@ucy.ac.cy}}}

\affil[1]{{\small In Silico Modelling Group, Department of Mechanical Engineering, University of Cyprus, Nicosia, Cyprus}}
\affil[2]{{\small New York University Abu Dhabi, Abu Dhabi, United Arab Emirates}}
\affil[3]{{\small Pontificia Universidad Católica de Valparaíso, Valparaíso, Chile}}
\affil[4]{{\small Weierstrass-Institut für Angewandte Analysis und Stochastik, Berlin, Germany}}
\affil[5]{{\small Charité-Universitätsmedizin Berlin, Campus Charité Mitte, Berlin, Germany}}
\affil[6]{{\small Department of Medical Physics \& Biomedical Engineering, University College London, London, United Kingdom}}

\date{}

\providecommand{\keywords}[1]
{
	\small	
	\textbf{\textit{Keywords:}} #1
}

\newcommand{\Grad}{\operatorname{grad}}
\newcommand{\Div}{\operatorname{div}}

\newcommand{\idem}{\Tilde{\mathbf{I}}}
\newcommand{\nv}{\mathbf{n}}

\newcommand{\Stress}{\boldsymbol{\sigma}}
\newcommand{\Strain}{\mathbf{e}}

\newtheorem{remark}{Remark}

\begin{document}

\maketitle

{
\section*{Abstract}
Accurate numerical modelling of wave propagation in brain magnetic resonance elastography (MRE) requires an appropriate representation of the mechanical interaction between viscoelastic brain tissue and cerebrospinal fluid (CSF). In this work, a monolithic mixed displacement--pressure $u–p$ finite element formulation is developed for the time-harmonic wave propagation in coupled viscoelastic solid–fluid domains in the frequency domain. The brain parenchyma is modelled as a linear viscoelastic solid, while the ventricular CSF is represented by an incompressible fluid governed by the unsteady Stokes equations. A residual-based stabilization derived from the Variational Multiscale Method (VMS) is introduced in the fluid subdomain to accommodate computationally efficient equal-order $P_1/P_1$ discretizations. The proposed framework is verified through manufactured--solution benchmarks involving individual solid and fluid problems and a coupled configuration with an embedded fluid inclusion. The results demonstrate the convergence and a substantial reduction in computational time compared with the Taylor--Hood $P_2/P_1$ formulation. Subsequently, the framework is applied to an anatomical brain model considering excitation frequencies in the range $15-45$~Hz. Comparisons with a simplified model employing uniform traction on the ventricular wall reveal differences in the spatial distribution of tissue displacement and highlight the role of the non--uniform CSF pressure and the kinematic constraints associated with fluid incompressibility. The findings of the present work demonstrate the importance of explicitly taking into account the effect of CSF for the investigation of viscoelastic wave propagation in brain MRE.
\\
\\
\keywords{Cerebrospinal Fluid, Magnetic Resonance Elastography, Brain, Viscoelasticity, Stokes flow, Mixed Finite Element Method, Variational Multiscale Method}
}

\section{Introduction}

Magnetic resonance elastography (MRE) is a non-invasive imaging technique that quantifies the mechanical properties of soft biological tissues in vivo \cite{muthupillai1995,hirsch2017book}. In MRE, low-frequency harmonic vibrations are introduced into the body by an external actuator, while the resulting steady-state displacement field is measured using motion-sensitive phase-contrast MRI. Then, the viscoelastic parameters of the tissue, such as the complex shear modulus, are recovered by solving an inverse problem based on a model of time-harmonic wave propagation \cite{hirsch2017book,manduca2021}. Since the mechanical properties reflect the microstructure and composition of tissue, MRE provides information complementary to conventional MRI contrasts.

One of the most active application areas of MRE is the brain \cite{hiscox2016,murphy2019}. Cerebral MRE has shown that brain viscoelasticity changes with age \cite{sack2009} and is sensitive to neurological disorders such as multiple sclerosis \cite{wuerfel2010}, Alzheimer's disease \cite{murphy2011} and normal pressure hydrocephalus \cite{streitberger2011}. However, the reported changes are often only a few percent of the healthy values. Therefore, the accuracy of the reconstructed parameters is critical for the clinical use of brain MRE. Most inversion methods assume a locally homogeneous, isotropic viscoelastic medium, an assumption that is violated near anatomical boundaries. Simulations have shown that cranial structures such as the falx cerebri, the tentorium cerebelli and the ventricles cause wave reflection and interference that can bias the reconstructed moduli \cite{mcgrath2016}.

Among the anatomical features that influence brain MRE, cerebrospinal fluid (CSF) requires particular consideration. CSF occupies the cerebral ventricles and the subarachnoid space, creating fluid--solid interfaces with the surrounding brain parenchyma. Because fluid and tissue exhibit distinct mechanical responses, their interaction can affect the displacement field in the vicinity of these interfaces. Experimental work using inversion--recovery MRE has demonstrated that suppressing the fluid signal changes the apparent ventricular boundaries and improves stiffness quantification near fluid--solid interfaces \cite{lilaj2021}. Related approaches have investigated the separation of fluid and solid wave contributions in magnetic resonance poroelastography \cite{perrinez2009,lilaj2020}. However, the latter concerns fluid contributions within a porous tissue representation and should be distinguished from the explicit treatment of the macroscopic CSF--filled ventricular cavities.

From a mechanical perspective, representing the ventricular CSF as a distinct fluid domain introduces more than a pressure load on the adjacent tissue. The fluid pressure is generally non--uniform, and the surrounding interface must satisfy both kinematic compatibility and traction balance. Moreover, for an enclosed incompressible fluid with an impermeable interface and no sources or sinks, the net normal displacement of the ventricular boundary vanishes in the linearized time--harmonic setting, expressing conservation of the enclosed fluid volume. Replacing the fluid by a prescribed traction on the ventricular wall does not, by itself, enforce this constraint. Consequently, the choice of the CSF representation may alter the computed wave field even when the imposed traction has a magnitude comparable to the fluid pressure.

Computational models provide a means to examine how material properties, anatomical structures, and excitation conditions influence MRE measurements. Finite element models have been used to investigate the effects of cranial anatomical structures on wave propagation and modulus reconstruction \cite{mcgrath2016}, as well as the sensitivity of brain MRE to the method of vibration delivery \cite{mcgrath2017delivery}. The role of constitutive assumptions has also been investigated through viscoelastic and poroelastic MRE models \cite{McGarry2015,perrinez2009,galarce2023}, while mathematical analyses have addressed the nearly incompressible viscoelastic equations underlying MRE \cite{jiang2011}. More recently, mixed modelling approaches have supported the reconstruction of displacement and pressure from MRE images in an in silico brain setting \cite{Fovargue2018Review,Fovargue2018Reconstruction,galarce2023}.

Beyond their role in understanding wave propagation and its dependence on modelling assumptions, accurate forward numerical solvers can also serve as tools for generating physically consistent synthetic data. Such data can support the development and assessment of reduced-order models \cite{galarce2023,Galarce2021} and physics-informed computational approaches \cite{rodopoulos2025vpinn}. These complementary applications highlight the importance of reliable numerical formulations capable of representing the relevant mechanical interactions. In this context, the present work develops a coupled finite element framework that explicitly resolves the mechanical response of brain tissue and ventricular CSF.

A central numerical difficulty arises from the incompressibility constraint in the fluid and, potentially, from the nearly incompressible behavior of brain tissue. For standard mixed Galerkin discretizations of incompressible problems, the choice of the discrete spaces is governed by the inf--sup condition, for which the Taylor–Hood $P_2/P_1$ pair is a classical choice \cite{Boffi2013,taylorhood1973}. Equal--order $P_1/P_1$ interpolation functions are computationally attractive for large three--dimensional meshes. However, this interpolation pair is not inf--sup stable for the incompressible Stokes problem without additional stabilization. Stabilized finite element formulations have therefore been developed to accommodate equal--order approximations, including pressure--stabilized methods \cite{hughes1986} and approaches based on the Variational Multiscale (VMS) framework introduced by Hughes \cite{Hughes1995,hughes1998}. In VMS methods, the effect of unresolved solution scales is incorporated into the resolved finite element equations. Related developments include orthogonal subscale formulations for incompressible flows \cite{codina2000,codina2002} and stabilized mixed formulations for nearly incompressible elasticity and fluid flow \cite{masud2005,Masud2009}. Subgrid--scale ideas have also been investigated for the treatment of transmission conditions between fluid and solid domains \cite{codinaBaiges2011}. These studies provide a foundation for stabilized mixed discretizations of the coupled brain--CSF problem.

Motivated by these considerations, the present work focuses on the development and investigation of a monolithic mixed displacement--pressure $u-p$ finite element framework for time--harmonic wave propagation in coupled viscoelastic solid and incompressible fluid domains. The brain parenchyma is described by linear viscoelasticity, while the ventricular CSF is governed by the unsteady Stokes equations, expressed in the frequency domain using displacement as a common kinematic variable. The formulation enforces displacement and traction continuity at the fluid--solid interface while retaining separate pressure fields in the two subdomains. A residual-based VMS stabilization, following the works \cite{masud2005,Masud2009}, is introduced in the fluid equations to accommodate an equal--order $P_1/P_1$ approximation. The numerical methodology is first verified using manufactured solutions for single--domain solid and fluid problems and a coupled solid--fluid configuration, with accuracy and computational cost compared against the Taylor--Hood $P_2/P_1$ formulation. Subsequently, the proposed numerical framework is implemented to an anatomical brain model considering multiple MRE excitation frequencies. The role of ventricular CSF is examined by comparing the coupled response with that of a simplified model in which the fluid is removed and a spatially uniform normal traction is prescribed on the ventricular walls.

The remainder of the paper is organized as follows. Section~2 presents the governing boundary value problem, the monolithic mixed finite element formulation as well as the VMS stabilization. Section~3 describes the verification benchmarks and the application to the anatomical brain--CSF model. Finally, the concluding section summarizes the main findings and outlines directions for future work.

\section{Methods}

\subsection{Wave propagation in the brain/CSF configuration}
\label{Sec:StrongForm}

\begin{figure}[H]
\centering
\includegraphics[scale=.35]{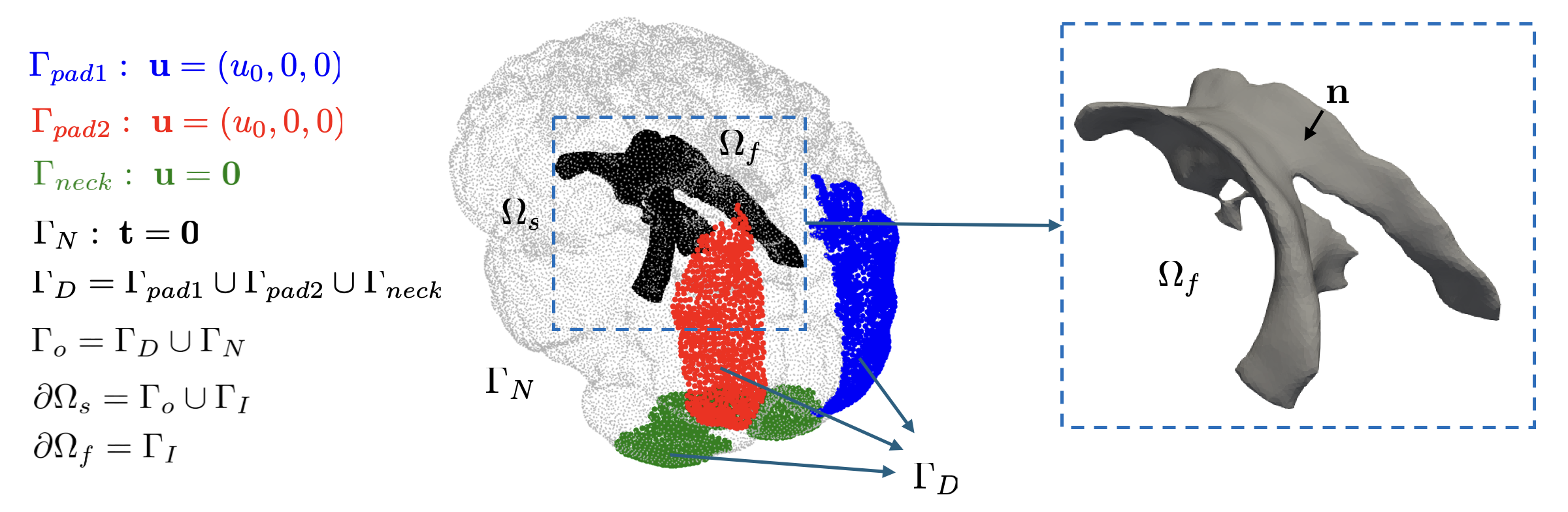} 
\caption{The brain--CSF model. The brain tissue is denoted by $\Omega_{s}$, while the cerebrospinal fluid with $\Omega_{f}$. Let $\mathbf{n}$ denote the unit normal vector pointing from $\Omega_{s}$ to $\Omega_{f}$. The outer boundary $\Gamma_{o} = \Gamma_{D} \cup \Gamma_{N}$ splits into the Dirichlet and Neumann part. The red and blue coloured segments of $\Gamma_{D}$ represent the areas where the harmonic load of the pads is applied, whereas the green coloured segment denotes the boundary where the displacement is fixed.} 
\label{fig:bvp}
\end{figure}

The brain model is considered as a bounded domain $\Omega \subset \mathbb{R}^{3}$ partitioned into a solid region $\Omega_{s}$, representing the brain parenchyma, and a fluid region $\Omega_{f}$, representing the cerebrospinal fluid (CSF) in the ventricles, such that $\overline{\Omega} = \overline{\Omega}_{s} \cup \overline{\Omega}_{f}$ and $\Omega_{s} \cap \Omega_{f} = \emptyset$. The fluid region is enclosed by the solid, so the solid-fluid interface is denoted by $\Gamma_{I} = \partial \Omega_{f}$ with $\nv$ denoting the unit normal vector pointing from $\Omega_{s}$ into $\Omega_{f}$, whereas $\partial \Omega_{s} = \Gamma_{o} \cup \Gamma_{I}$ is the boundary of the brain, as shown in Fig \ref{fig:bvp}. The outer boundary consists from a Dirichlet part and a Neumann part, such that: $\Gamma_{o} = \Gamma_{D} \cup \Gamma_{N}$, with $\Gamma_{D} \cap \Gamma_{N} = \emptyset$.

In MRE, the tissue is driven by a harmonic excitation of angular frequency $\omega = 2\pi f$ with the steady-state displacement field being the measured quantity. Therefore, time-harmonic fields of the form $\mathbf{u}(\mathbf{x},t) = \Re\{\hat{\mathbf{u}}(\mathbf{x},\omega)\, e^{i\omega t}\}$ are assumed and the analysis is carried out in the frequency domain using the complex amplitudes, omitting the hat symbol for brevity. Under this assumption, the time-domain boundary value problem is transformed into the frequency domain via the procedure detailed in Appendix A. To this point, the brain parenchyma is modelled as a homogeneous and isotropic medium, following linear viscoelastic behaviour with complex shear modulus $\mu_{s} = \mu_{s}' + i\mu_{s}''$, where the storage modulus $\mu_{s}'$ and the loss modulus $\mu_{s}''$ describe the elastic and dissipative effects, respectively.
The CSF is modelled as an incompressible Newtonian fluid with dynamic viscosity $\eta_{f}$ and density $\rho_{f}$. Given the small amplitude of the MRE excitation, convective effects are neglected and the fluid motion is governed by the unsteady Stokes equations. In order to employ a common kinematic variable in both subdomains, the fluid equations are expressed in terms of the displacement vector $\mathbf{u}_{f}$, which, in the frequency domain, is related to the velocity via $\partial_{t} \mapsto i\omega$ (as stated in Appendix A). Therefore, based on a mixed formulation for viscoelasticity, the complex displacement $\mathbf{u}_{s}: \Omega_{s} \rightarrow \mathbb{C}^{3}$, $\mathbf{u}_{f}: \Omega_{f} \rightarrow \mathbb{C}^{3}$ and pressure functions $p_{s}: \Omega_{s} \rightarrow \mathbb{C}$, $p_{f}: \Omega_{f} \rightarrow \mathbb{C}$  satisfy the following set governing partial differential equations:
\begin{equation}
\begin{aligned}
2\mu_{s} \Div \mathbf{e} (\mathbf{u}_{s}) - \nabla p_{s} + \rho_{s} \omega^{2} \mathbf{u}_{s} &= \mathbf{0} && \text{in } \Omega_{s} 
\\
\Div\mathbf{u}_{s} + \frac{1}{\lambda_{s}}p_{s}&= 0 && \text{in } \Omega_{s}
\\
2 i\omega\eta_{f}\, \Div  \mathbf{e}(\mathbf{u}_{f}) - \nabla p_{f} + \rho_{f} \omega^{2} \mathbf{u}_{f}  &= \mathbf{0} && \text{in } \Omega_{f}
\\
\Div \mathbf{u}_{f} &= 0 && \text{in } \Omega_{f}.
\end{aligned}
\label{Eq:SolidFluid}
\end{equation}
where $\Div = \nabla \cdot$, $\mathbf{e} = \tfrac{1}{2}\left(\nabla \mathbf{u} + (\nabla \mathbf{u})^{T}\right)$ is the infinitesimal strain tensor, $\rho_{s}$ the tissue density and $\lambda_{s}$ the first complex Lame's parameter. The second equation of Eq. \eqref{Eq:SolidFluid}, shows that as we approach the incompressible limit ($\nu = 0.5$), the parameter $\lambda_{s} \rightarrow \infty$ and $\frac{1}{\lambda_{s}} \rightarrow 0$, making the pressure $p_{s}$ acting as a Lagrange multiplier in the occurred saddle point problem.
Since $\omega \neq 0$, the incompressibility of the velocity is equivalent to that of the displacement.
In order to express the governing equations of both subdomains in a unified form, the subdomain-wise coefficients $\kappa_{j}$ and $\gamma_{j}$, such that $\kappa_{s} = \mu_{s} = \mu_{s}' + i \mu_{s}''$, $\kappa_{f} = i\omega\eta_{f}$, $\gamma_{s} = \frac{1}{\lambda_{s}}$ and $\gamma_{f} = 0$ with $j \in \{s,f\}$ are defined and the following set of equations is obtained:
\begin{equation}
\begin{aligned}
2\kappa_{j}\, \Div  \mathbf{e}(\mathbf{u}_{j}) - \nabla p_{j} + \rho_{j} \omega^{2} \mathbf{u}_{j}  &= \mathbf{0} && \text{in } \Omega_{j}, \\
\Div \mathbf{u}_{j} + \gamma_{j}p_{j}&= 0 && \text{in } \Omega_{j}.
\end{aligned}
\label{Eq:Unified}
\end{equation}

The continuity of displacement $\mathbf{u}$ and traction $\mathbf{t}(\mathbf{u},p)$ are imposed on the fluid--solid interface $\Gamma_{I}$, i.e:
\begin{equation}
\mathbf{u}_{s} = \mathbf{u}_{f}, \qquad \mathbf{t}(\mathbf{u}_{s},p_{s}) + \mathbf{t}(\mathbf{u}_{f},p_{f}) = \mathbf{0} \qquad \text{on } \Gamma_{I}.
\label{Eq:Interface}
\end{equation}
where $\mathbf{t}(\mathbf{u}_{j},p_{j}) = \boldsymbol{\sigma}(\mathbf{u}_{j},p_{j}) \cdot \nv_{j}$ and $\boldsymbol{\sigma}(\mathbf{u}_{j},p_{j}) = -p_{j} \idem + 2 \kappa_{j} \mathbf{e}(\mathbf{u}_{j})$ denotes the Cauchy stress tensor. The first condition of Eq. \eqref{Eq:Interface} corresponds to the frequency domain version of the no-slip condition for the fluid. On the outer boundary $\Gamma_{o} = \Gamma_{D} \cup \Gamma_{N}$, the following boundary conditions are prescribed:
\begin{equation}
\left.\begin{matrix}
\mathbf{u}_{s} = \mathbf{g} \quad \text{on } \Gamma_{pad1} \cup \Gamma_{pad2}
\\ 
\mathbf{u}_{s} = \mathbf{0} \quad \text{on } \Gamma_{neck}
\end{matrix}\right\}
\quad \Gamma_{D} = \Gamma_{pad1} \cup \Gamma_{pad2} \cup \Gamma_{neck}, 
\qquad \mathbf{t}_{s}(\mathbf{u}_{s},p_{s}) = \mathbf{h} \quad \text{on } \Gamma_{N},
\label{Eq:BCs}
\end{equation}
where $\mathbf{g}$ models the actuator excitation and, in the applications below, $\mathbf{h} = \mathbf{0}$.

\begin{remark}
\label{rem:pcont}
It should be noted that the continuity of traction does not imply that pressure is continuous on $\Gamma_{I}$. Taking the normal part of the traction vector $\mathbf{t}_{n}(\mathbf{u},p) = -p + 2 \kappa\, e_{nn}(\mathbf{u})$, as shown in the Appendix A that pressure continuity holds only if:
\begin{equation}
\frac{\kappa_{s}}{\lambda_{s}}p_{f} = (\kappa_{s}-\kappa_{f})e_{nn}(\mathbf{u}_{f})
\label{Eq:pcontin}
\end{equation}
where $e_{nn}(\mathbf{u}) = \nv \cdot \nabla \mathbf{u} \cdot \nv$. This is reflected in the choice of the pressure space, as it is reported in Section \ref{Sec:WeakForm}.
\end{remark}

The present formulation should be distinguished from classical Biot poroelasticity \cite{Carcamo2024,galarce2023}, in which the solid skeleton and interstitial fluid are represented as overlapping continua within a porous medium, where their interaction described through constitutive and inertial coupling parameters. In the present study, the brain parenchyma is instead treated as a single-phase viscoelastic continuum, while the cerebrospinal fluid occupying the ventricles is explicitly modeled as a separate incompressible fluid. The interaction between these domains is described through kinematic compatibility and traction equilibrium on their shared interface boundary. Consequently, fluid transport through the tissue microstructure as well as the associated poroelastic coupling mechanisms are not considered. Therefore, the present approach focuses on the macroscopic brain–CSF interactions and is complementary to poroelastic models of brain tissue \cite{lilaj2020,McGarry2015,perrinez2009}.

\subsection{Mixed $u-p$ Finite Element formulation}
\label{Sec:WeakForm}

In order to derive the finite element formulation, the weak formulation of the boundary value problem \eqref{Eq:Unified}-\eqref{Eq:BCs} should be defined. To this point, the following function spaces are introduced:
\begin{equation}
\begin{aligned}
\mathcal{U} &= \big\{ (\mathbf{u}_{s},\mathbf{u}_{f}) \in H^{1}(\Omega_{s}; \mathbb{C}^3) \times H^{1}(\Omega_{f}; \mathbb{C}^3) \;\big|\; \mathbf{u}_{s} = \mathbf{g} \text{ on } \Gamma_{D},\; \mathbf{u}_{s} = \mathbf{u}_{f} \text{ on } \Gamma_{I} \big\}, \\
\mathcal{U}_{0} &= \big\{ (\mathbf{w}_{s},\mathbf{w}_{f}) \in H^{1}(\Omega_{s}; \mathbb{C}^3) \times H^{1}(\Omega_{f}; \mathbb{C}^3) \;\big|\; \mathbf{w}_{s} = \mathbf{0} \text{ on } \Gamma_{D},\; \mathbf{w}_{s} = \mathbf{w}_{f} \text{ on } \Gamma_{I} \big\}, \\
\mathcal{Q} &= L^{2}(\Omega_{s}; \mathbb{C}) \times L^{2}(\Omega_{f}; \mathbb{C})
\end{aligned}
\label{Eq:Spaces}
\end{equation}
where $L^{2}(\Omega_{j})$ is the space of the square-integrable functions on $\Omega_{j}$, while $H^{1}(\Omega_{j})$ denotes the space of the square-integrable functions with square-integrable first derivatives on $\Omega_{j}$. It should be noted that pressure space imposes no continuity across $\Gamma_{I}$, in agreement with Remark \ref{rem:pcont}. 
Multiplying~\eqref{Eq:Unified} by test functions $\mathbf{w}_{j} \in \mathcal{U}_{0}$, $\upsilon_{j} \in Q$, integrating by parts over for each $\Omega_{j}$, the weak formulation coupled solid--fluid problem is:

\medskip
\noindent Find $(\mathbf{u}_{s},\mathbf{u}_{f},p_{s},p_{f}) \in \mathcal{U}\times\mathcal{Q}$ such that
\begin{equation}
\begin{aligned}
\sum_{j\in\{s,f\}} \Big[ 
&\int_{\Omega_{j}} 2\kappa_{j}\, \mathbf{e}(\mathbf{w}_{j}) : \mathbf{e}(\mathbf{u}_{j})\,d\Omega -
\int_{\Omega_{j}} \rho_{j}\, \omega^2 \, \mathbf{w}_{j} \cdot \mathbf{u}_{j}\,d\Omega -
\int_{\Omega_{j}} (\Div \mathbf{w}_{j})\, p_{j}\,d\Omega - 
\\
&\int_{\Omega_{j}} (\Div \mathbf{u}_{j})\, \upsilon_{j}\,d\Omega - 
\int_{\Omega_{j}} \gamma_{j} \, \upsilon_{j} \, p_{j}\,d\Omega
\Big] = \int_{\Gamma_{N}} \mathbf{w}_{s}\cdot\mathbf{h}\, d\Gamma
\end{aligned}
\label{Eq:Weak}
\end{equation}
for all $(\mathbf{w}_{s},\mathbf{w}_{f},\upsilon_{s},\upsilon_{f}) \in \mathcal{U}_{0}\times\mathcal{Q}$
\medskip

\noindent
The summation of the integration by parts result of Eq. \eqref{Eq:Weak}, reveals the following surface integrations:
\begin{equation}
\begin{aligned}
&\int_{\partial \Omega_{s}} \mathbf{w}_{s} \cdot \big( \Stress(\mathbf{u}_{s},p_{s}) \cdot \nv_{s} \big) \,d\Gamma + 
\int_{\partial \Omega_{f}} \mathbf{w}_{f} \cdot \big( \Stress(\mathbf{u}_{f},p_{f}) \cdot \nv_{f} \big) \,d\Gamma = 
\\
&\int_{\Gamma_{I}} \mathbf{w} \cdot \big( \Stress_{s} - \Stress_{f}  \big)\cdot \nv \,d\Gamma +
\int_{\Gamma_{D}} \mathbf{w}_s \cdot \big( \Stress_{s} \cdot \nv \big) \,d\Gamma + 
\int_{\Gamma_{N}} \mathbf{w}_{s} \cdot \mathbf{h} \,d\Gamma =
\\
&\int_{\Gamma_{N}} \mathbf{w}_{s} \cdot \mathbf{h} \,d\Gamma
\end{aligned}
\label{Eq:Surfinteg}
\end{equation}
considering that $\partial \Omega_{s} = \Gamma_{I} \cup \Gamma_{N} \cup \Gamma_{D}$ and that the integral $\int_{\Gamma_{I}} \mathbf{w} \cdot \big( \Stress_{s} - \Stress_{f}  \big)\cdot \nv \,d\Gamma$ vanishes as the dynamic interface condition of Eq. \eqref{Eq:Interface} is satisfied. Thus, the continuity of traction as well the Neumann boundary condition \eqref{Eq:BCs} are imposed naturally (weakly) by the monolithic formulation. 

\paragraph{Discretization.} Let $\mathcal{T}_{h}$ be a conforming tetrahedral mesh of $\Omega$ that is fitted to the interface such that every element $E \in \mathcal{T}_{h}$ lies entirely in either $\Omega_{s}$ or $\Omega_{f}$, and $\Gamma_{I}$ is a union of element faces. We denote by $\mathcal{T}_{h}(\Omega_{j})$ the elements in $\Omega_{j}$. The discrete spaces are
\begin{equation}
\begin{aligned}
\mathcal{U}_{h} &= \big\{ \mathbf{u}_{h} \in \mathcal{U}  \;\big|\; \mathbf{u}_{h}|_{E} \in \mathcal{P}^{k}(E)^{3},\; E \in \mathcal{T}_{h}(\Omega) \big\}, \\
\mathcal{Q}_{h} &= \big\{ (\upsilon_{s,h},\upsilon_{f,h}) \in \mathcal{Q} \;\big|\; \upsilon_{j,h} \in C^{0}(\overline{\Omega}_{j}),\; \upsilon_{j,h}|_{E} \in \mathcal{P}^{1}(E),\; E \in \mathcal{T}_{h}(\Omega_{j}),\; j\in\{s,f\} \big\},
\end{aligned}
\label{Eq:DiscreteSpaces}
\end{equation}
where $\mathcal{P}^{k}(E)$ denotes polynomials of degree at most $k$ on $E$. 

The displacement and pressure fields are approximated using standard real-valued finite element shape functions, while the corresponding nodal unknowns are complex-valued. Within each element $E$, the discrete fields are expressed as: $\mathbf{u}_{h} = \sum_{l}N^{l}(\boldsymbol{\xi}) \hat{\mathbf{u}}^l$ and $p_{h} = \sum_{l}M^{l}(\boldsymbol{\xi}) \hat{p}^l$. Here, $N^l$ and $M^l$ denote the displacement and pressure interpolation functions, respectively, while $\hat{\mathbf{u}}^l \in \mathbb{C}^3$ and $\hat{p}^l \in \mathbb{C}$ are the corresponding nodal degrees of freedom in $E$. The complex-valued nodal coefficients represent the amplitudes and phases of the time-harmonic fields. The same interpolation procedure is employed in both subdomains, with separate pressure degrees of freedom assigned at the fluid–solid interface.

The displacement is continuous over the whole domain $\Omega$, so the kinematic condition holds exactly. The pressure is continuous within each subdomain but discontinuous across $\Gamma_{I}$. In the implementation, this is achieved by assigning two pressure degrees of freedom to every node on $\Gamma_{I}$: one used by the elements of $\mathcal{T}_{h}(\Omega_{s})$ and one used by the elements of $\mathcal{T}_{h}(\Omega_{f})$. Displacement degrees of freedom on $\Gamma_{I}$ remain shared. A continuous pressure across $\Gamma_{I}$ would impose the additional constraint $p_{s} = p_{f}$, which contradicts Remark \ref{rem:pcont} and limits the accuracy of the approximation near the interface.


In the present work, a stabilization based on Variational Multiscale Method (VMS) \cite{hughes1986,Masud2009} is employed in the fluid subdomain to accommodate equal-order interpolation, as detailed in Section 2.3 and Appendix B. The numerical accuracy, convergence behaviour and computational performance of the resulting coupled formulation are assessed through the verification problems presented in Section 3. A rigorous mathematical analysis of the well-posedness and stability of the complete coupled formulation constitutes a direction for future research.

\subsection{Variational Multiscale Method}
\label{Sec:vms}

The finite element formulation developed in the present work employs the Variational Multiscale (VMS) framework introduced by Hughes \cite{Hughes1995,hughes1998}, following the approach developed by Masud and co-workers for fluid and solid mechanics problems \cite{Masud2009,masud2005}. Several stabilization strategies have been proposed for mixed finite element formulations, including pressure-stabilized Petrov–Galerkin methods \cite{hughes1986} and orthogonal subscale approaches \cite{codina2000,codina2002}. The VMS framework provides a systematic approach for constructing residual-based stabilization methods by accounting for the influence of unresolved solution scales on the finite element approximation. Compared with conventional pressure-gradient stabilization approaches, the present VMS formulation approximates the unresolved displacement scales using element-level bubble functions, which are subsequently eliminated locally. This procedure leads to a residual-based stabilization term whose parameters are determined from the element geometry, material properties and excitation frequency, without requiring the selection of an independently tuned stabilization coefficient. This makes the approach particularly suitable for the present time-harmonic formulation, involving complex-valued coefficients and equal-order displacement–pressure interpolation.

The displacement and the corresponding test function are decomposed into a coarse (resolved) part and a fine (unresolved) part:
\begin{equation}
\mathbf{u} = \mathbf{u}_{h} + \mathbf{u}', \qquad \mathbf{w} = \mathbf{w}_{h} + \mathbf{w}',
\end{equation}
while the pressure is represented only at the coarse scale, $p = p_{h}$, $\upsilon = \upsilon_{h}$. The fine scale parts $\mathbf{u}'$, $\mathbf{w}'$ are defined in each element $E$ and assumed to vanish on $\partial E$. Following the Variational Multiscale procedure, given in detail in Appendix B, the fine scale displacement is calculated locally inside an element $E$ as:
\begin{equation}
\mathbf{u}' = \boldsymbol{\tau} \cdot \mathcal{R}(\mathbf{u}_{h},p_{h}),
\qquad
\boldsymbol{\tau}(\mathbf{x}) = b_{E}(\mathbf{x}) \left( \int_{E} b_{E}\, d\Omega \right) \mathbf{K}_{E}^{-1},
\label{Eq:FineSol1}
\end{equation}
with
\begin{equation}
\mathbf{K}_{E} = \int_{E} \kappa \left( |\nabla b_{E}|^{2}\,\mathbf{I} + \nabla b_{E}\otimes\nabla b_{E} \right) d\Omega - \rho\omega^{2}\int_{E} b_{E}^{2}\, d\Omega\; \mathbf{I}.
\label{Eq:Tau1}
\end{equation}
where $\mathcal{R}_{f}(\mathbf{u}_{h},p_{h}) = 2 \kappa_{f} \, \Div \mathbf{e}(\mathbf{u}_{f}) - \nabla p_{h} + \rho_{f} \, \omega^{2}\mathbf{u}_{f}$. The stabilization tensor $\boldsymbol{\tau}$ is complex-valued, symmetric, and computed element by element from the material parameters and the element geometry. According to the process in Appendix B, we obtain the following stabilized finite element formulation:

\medskip
\noindent Find $(\mathbf{u}_{h},p_{h}) \in \mathcal{U}_{h}\times\mathcal{Q}_{h}$ such that
\begin{equation}
\begin{aligned}
\sum_{j\in\{s,f\}} \Big[ 
&\int_{\Omega_{j}} 2\kappa_{j}\, \mathbf{e}(\mathbf{w}_{j}) : \mathbf{e}(\mathbf{u}_{j})\,d\Omega -
\int_{\Omega_{j}} \rho_{j}\, \omega^2 \, \mathbf{w}_{j} \cdot \mathbf{u}_{j}\,d\Omega -
\int_{\Omega_{j}} (\Div \mathbf{w}_{j})\, p_{j}\,d\Omega - 
\\
&\int_{\Omega_{j}} (\Div \mathbf{u}_{j})\, \upsilon_{j}\,d\Omega - 
\int_{\Omega_{j}} \gamma_{j} \, \upsilon_{j} \, p_{j}\,d\Omega
\Big] - 
\sum_{E\in\mathcal{T}_{h}^{f}} 
\int_{E}
 \boldsymbol{\tau} \cdot \mathcal{R}_{f}(\mathbf{w}_{h},\upsilon_{h}) \cdot \mathcal{R}_{f}(\mathbf{u}_{h},p_{h}) \, d\Omega = \\
 &\int_{\Gamma_{N}} \mathbf{w}_{s}\cdot\mathbf{h}\, d\Gamma    
\end{aligned}
\label{Eq:VMS}
\end{equation}
for all $(\mathbf{w}_{h},\upsilon_{h}) \in \mathcal{U}_{0,h}\times\mathcal{Q}_{h}$.
\medskip

\noindent
The stabilization term is symmetric and vanishes for the exact solution of the homogeneous governing equations, preserving the consistency of the formulation. Its discretization introduces element-local contributions to the mixed finite element system, including a pressure-gradient term intended to address the numerical difficulties associated with equal-order $P_1/P_1$ interpolation in the incompressible fluid subdomain. The stabilization integrals involve no inter-element face terms and are therefore compatible with the adopted pressure space $Q_h$, which allows pressure discontinuities across $\Gamma_I$.

The proposed formulation has been implemented in an in-house finite element code developed using C++ language. The implementation employs the Eigen \cite{Guennebaud2010} and PETSc \cite{Balay2025} libraries for linear algebra operations and the solution of the resulting sparse linear system. The discretized systems are solved using the Generalized Minimal Residual (GMRES) \cite{saad2003iterative} iterative method, equipped with an incomplete LU (ILU) preconditioner. Shared-memory parallelism is supported through OpenMP \cite{dagum1998openmp} for multithreaded computations.

\FloatBarrier
\section{Results}

The proposed stabilized mixed FE formulation is assessed in two stages. Firstly, its accuracy and efficiency are verified with manufactured solutions of proceeding from two single--domain problems, isolating the viscoelastic solid and the Stokes fluid, to a coupled problem with an embedded fluid inclusion. In each case, the VMS-stabilized equal-order $P_1/P_1$ formulation is compared with the inf--sup stable Taylor--Hood $P_2/P_1$ pair. Secondly, the framework is applied to the time-harmonic wave propagation in an anatomical brain model including the CSF inside the ventricular system. The effect of the presence of the CSF investigated comparing with a simplified case where a uniform pressure is applied on the ventricle boundary as a traction boundary condition.

For the verification tests, the following relative error measures are used:
\begin{equation}
\begin{aligned}
e_{L^{2}(\Omega)}(\mathbf{u}) = \frac{\| \mathbf{u} - \mathbf{u}_{h} \|_{L^2(\Omega)}}{\| \mathbf{u} \|_{L^2(\Omega)}}, \qquad 
e_{L^{2}(\Omega)}(p) = \frac{\| p - p_h \|_{L^2(\Omega)}}{\| p \|_{L^2(\Omega)}}, \qquad
e_{H^{1}(\Omega)}(\mathbf{u}) = \frac{\| \mathbf{u} - \mathbf{u}_{h} \|_{H^1(\Omega)}}{\| \mathbf{u} \|_{H^1(\Omega)}}
\\
\| \mathbf{u} \|_{L^2(\Omega)} = \Big( \int_{\Omega} \mathbf{u} \cdot \bar{\mathbf{u}} \,d\Omega \Big)^{1/2}, \qquad \| \mathbf{u} \|_{H^1(\Omega)} = \Big( \int_{\Omega} (\mathbf{u} \cdot \bar{\mathbf{u}}  + \nabla \mathbf{u} :  \overline{\nabla \mathbf{u}}) \,d\Omega \Big)^{1/2}
\end{aligned}
\label{Eq:errors}
\end{equation}
where the overbar denotes complex conjugation.

\subsection{Framework verification for single--domain cases}
\label{Sec:b1}

The first step of the accuracy assessment of the proposed framework concerns two single domain test cases. Considering the unit cube $\Omega = (0,1)^{3}$, the first case assumes a viscoelastic material satisfying the first two equations Eq. \eqref{Eq:SolidFluid} on every point, while the second case focuses on a fluid satisfying the Stokes equation of \eqref{Eq:SolidFluid}. The mixed FEM is implemented considering the triangulation $\mathcal{T}(\Omega)$ of tetrahedral elements of size $h \in \{1/5, 1/10, 1/20, 1/30, 1/40\}$. In each case, the manufactured
displacement is imposed as a Dirichlet condition on the entire boundary
$\partial\Omega$, and the corresponding body force, obtained by applying the
differential operator to the manufactured fields, is evaluated at the
quadrature points of each element.

\paragraph{Viscoelastic solid case.} We prescribe the following complex displacement field, while the pressure is obtained applying the divergence operator:
\begin{equation}
\begin{aligned}
\mathbf{u}(\mathbf{x}) &= (\sin \pi x_{1} \cos \pi x_{2}  \cos \pi x_{3}, 
                          \cos \pi x_{1} \sin \pi x_{2}  \cos \pi x_{3}, 
                          \cos \pi x_{1} \cos \pi x_{2}  \sin \pi x_{3})
\\
p(\mathbf{x}) &= -\lambda \Div \mathbf{u} = -3 \lambda  \pi \cos \pi x_{1} \cos \pi x_{2}  \cos \pi x_{3}
\end{aligned}
\label{Eq:ms1}
\end{equation}
where the pressure follows from the continuity relation
in~\eqref{Eq:SolidFluid}. Since $\Div \mathbf{u}\neq 0$, this case tests the compressible regime, in which the term $\lambda_s^{-1}p$ is active. For the analysis, the following material parameters are considered: $\mu_s = 1 + i \, 0.1$, $\nu = 0.45$, $\rho_s = 1$, $f = 1$.

\paragraph{Stokes flow case.} For the Stokes problem, we prescribe the following divergence free displacement field and pressure:
\begin{equation}
\begin{aligned}
\mathbf{u}(\mathbf{x}) &= (\sin \pi x_{1} \cos \pi x_{2}  \cos \pi x_{3}, 
                          -\cos \pi x_{1} \sin \pi x_{2}  \cos \pi x_{3}, 
                          0)
\\
p(\mathbf{x}) &=  \sin \pi x_{1} \sin \pi x_{2}  \sin \pi x_{3}
\end{aligned}
\label{Eq:ms2}
\end{equation}
given the following parameters: $\eta_f = 1$, $\rho_f = 1$, $f = 1$. Here, the incompressibility constraint holds exactly, so the problem has the full saddle-point structure. This is the setting in which equal-order elements require stabilization.

\begin{figure}[!htb]
\centering
\includegraphics[scale=.5]{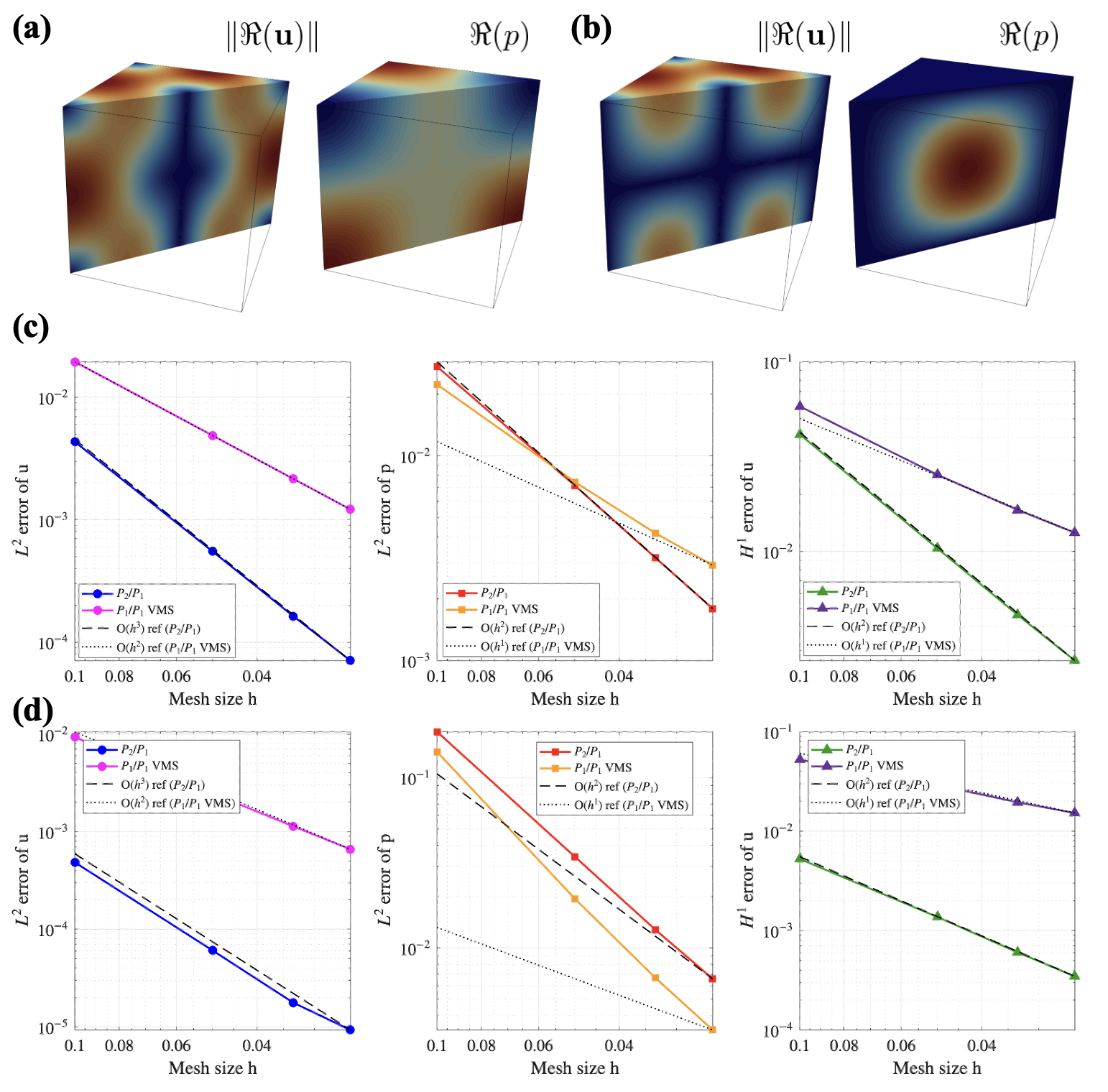} 
\caption{Verification of the finite element formulation using manufactured solutions on the unit cube $\Omega=(0,1)^3$. Numerical results for the real displacement magnitude $\|\Re(\mathbf{u})\| = \sqrt{\Re(\mathbf{u})\cdot \Re(\mathbf{u})}$ and real pressure part $\Re(p)$ for (a) viscoelastic solid case and (b) Stokes flow case. Mesh convergence study comparing the Taylor–Hood $P_2/P_1$ and the VMS-stabilized $P_1/P_1$ formulations in terms of errors norms of Eq. \eqref{Eq:errors} for the (c) viscoelastic solid and (d) Stokes flow.}
\label{fig:test1}
\end{figure}

The numerical solutions for the viscoelastic solid and Stokes flow cases are illustrated in Figs. \ref{fig:test1}a and \ref{fig:test1}b, respectively, while the corresponding mesh convergence analysis is presented in Figs. \ref{fig:test1}c and \ref{fig:test1}d. The accuracy of the proposed VMS-stabilized $P_1/P_1$ formulation is assessed against the inf–sup stable Taylor–Hood $P_2/P_1$ formulation, considering the error indicators of Eq. \eqref{Eq:errors}.
For the viscoelastic solid case, Fig. \ref{fig:test1} demonstrates a systematic reduction in the displacement and pressure errors under mesh refinement for both formulations. The Taylor–Hood formulation achieves the expected convergence rates of approximately $O(h^3)$ and $O(h^2)$ for the displacement in the $L^2$ and $H^1$ norms, respectively, while the pressure error decreases approximately as $O(h^2)$. The VMS-stabilized $P_1/P_1$ formulation exhibits approximately $O(h^2)$ convergence in the displacement $L^2$--norm and $O(h)$ convergence in the displacement $H^1$-norm, consistent with the approximation order of the linear finite elements. The pressure error also decreases with mesh refinement, approaching the expected first-order convergence rate on the finer meshes. These results verify the consistency and convergence behavior of the mixed formulation for a compressible viscoelastic solid with complex-valued material properties.

For the Stokes flow case, Fig. \ref{fig:test1}d illustrates the convergence behavior in the incompressible limit, where the pressure acts as a Lagrange multiplier enforcing the divergence-free constraint. The Taylor–Hood formulation retains approximately third-order convergence in the displacement $L^2$--norm and second-order convergence in the $H^1$--norm. Similarly, the VMS-stabilized $P_1/P_1$ formulation exhibits approximately second-order and first-order convergence in the corresponding norms. These results demonstrate that the stabilized equal-order formulation preserves the expected displacement approximation accuracy in the presence of the incompressibility constraint. Overall, the two verification cases confirm the expected convergence behaviour of the employed finite element formulations for both compressible viscoelastic and incompressible fluid problems, providing a basis for their subsequent application to the coupled fluid–solid system.

\FloatBarrier
\subsection{Framework verification on the coupled Elastic/Stokes case}
\label{Sec:b2}

\begin{figure}[!htb]
\centering
\includegraphics[scale=.5]{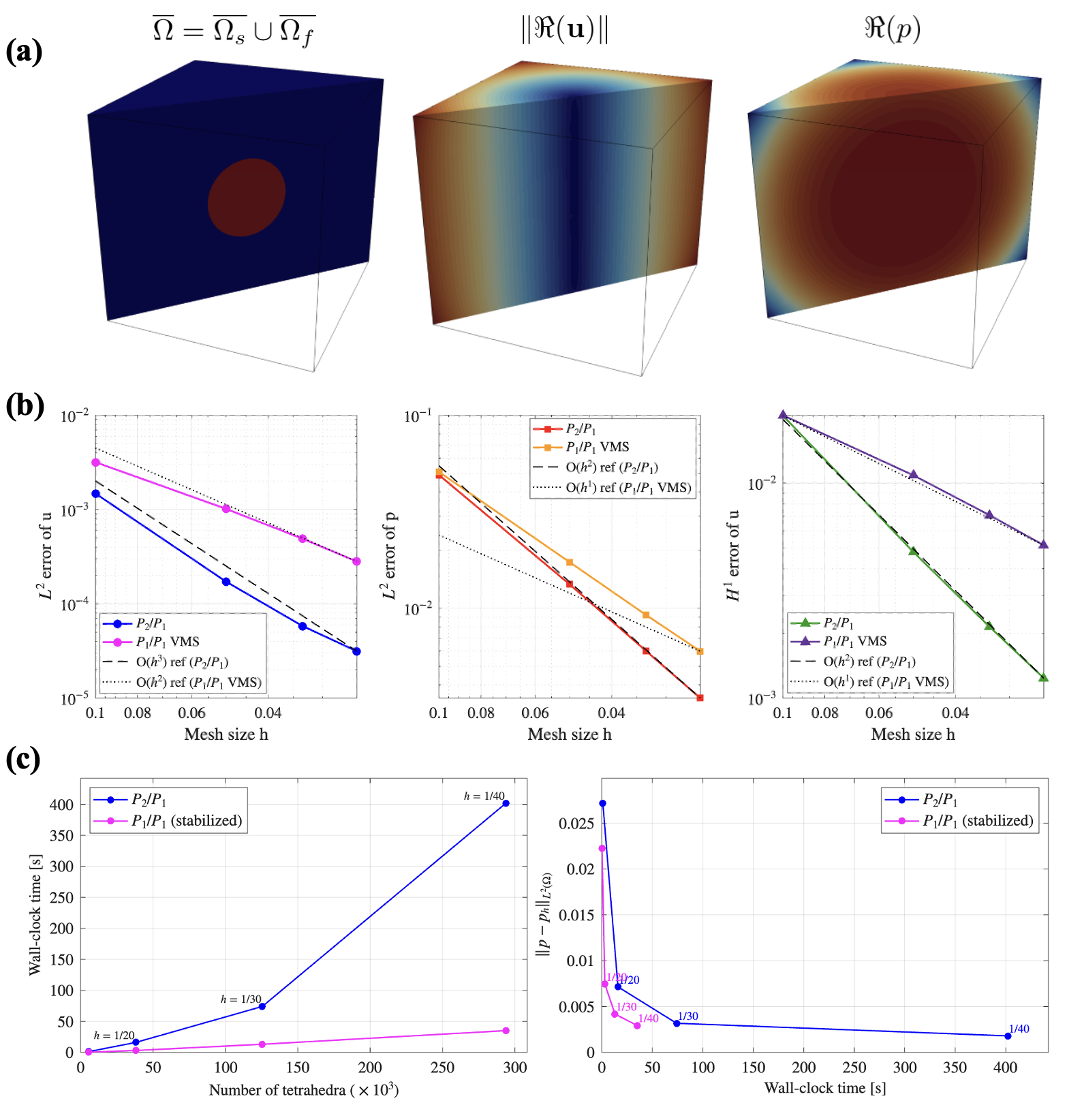} 
\caption{Verification of the coupled viscoelastic solid–Stokes formulation using a manufactured solution. (a) Computational domain consisting of a unit cube $\Omega=(0,1)^3$ with a spherical fluid inclusion, together with the numerical results for $\|\Re(\mathbf{u})\|$ and $\Re(p)$. (b) Mesh convergence study comparing the Taylor–Hood $P_2/P_1$ and VMS-stabilized $P_1/P_1$ formulations in terms of the displacement $L^2$- and $H^1$-errors. (c) Computational efficiency comparison showing the wall-clock time as a function of the number of tetrahedral elements (left) and the pressure $L^2$-error as a function of wall-clock time (right). }
\label{fig:test2}
\end{figure}

The coupled formulation is verified on the unit cube $\Omega = (0,1)^{3}$ containing a spherical fluid inclusion $\Omega_f$ of radius $r=0.2$, centered on $\mathbf{x}_c = (0.5,0.5,0.5)$, with the remaining domain of the cube occupied by the viscoelastic solid $\Omega_s$ (Fig.~\ref{fig:test2}a). The following manufactured solution that satisfies the kinematic and dynamic interface conditions~\eqref{Eq:Interface}, is considered:
\begin{equation}
\begin{aligned}
&\mathbf{u}_{f} = (-Y,X,0),
&p_{f} &= 0, &\qquad \text{in } \Omega_{f}
\\
&\mathbf{u}_{s} = \mathbf{u}_{f} + \phi^{2}\mathbf{q}, 
&p_{s} &= -\lambda_{s} \Div \mathbf{u}_{s}, &\qquad \text{in } \Omega_{s}
\end{aligned}
\label{Eq:ms3}
\end{equation}
where $(X,Y,Z) = \mathbf{x} - \mathbf{x}_{c}$, is $\mathbf{q}$ a vector field defined as $\mathbf{q}=(X,Y,Z)$ and $\phi$ denotes the function $\phi(r) = r^2 - a^2$, with $r = |\mathbf{x}-\mathbf{x}_{c}|$. The function $\phi(r)$ ensures that the displacement field is continuous (Eq.~\eqref{Eq:Interface}) on the interface $\Gamma_{I} = \{ \mathbf{x} \in \Omega \,|\, \phi(r) = 0\}$. The pressure of the solid part is calculated as:
\begin{equation}
p_s = - \lambda_s \Div \mathbf{u}_s = - \lambda_s (4\phi r^2 + 3\phi^2)
\label{Eq:ms1_ps}
\end{equation}
As $\Grad \mathbf{u}_{f} = -\hat{\mathbf{x}}_2 \otimes \hat{\mathbf{x}}_1 + \hat{\mathbf{x}}_1 \otimes \hat{\mathbf{x}}_2$, employing \eqref{Eq:ms1_ps}, the stress tensor for the solid and fluid domains take the form:
\begin{equation}
\begin{aligned}
&\Stress_f = \mathbf{0}
\\
&\Stress_s =  \lambda_s \Div \mathbf{u}_s + \mu_{s}\big(\Grad \mathbf{u}_s + (\Grad\mathbf{u}_s)^{T} \big) = 4 \mu_s \, \phi \, \mathbf{q} \otimes \mathbf{q} + 4 \lambda_s \phi \, r^2 \idem + (3\lambda_s + \mu_s)\phi^2\idem
\end{aligned}
\label{Eq:ms1_ss}
\end{equation}
Therefore, for all points on $\Gamma_{I}$ ($r=a$), according to \eqref{Eq:ms1_ps}--\eqref{Eq:ms1_ss}, $p_s = p_f = 0$ and $\Stress_{f} = \Stress_{s} = \mathbf{0}$ which also lead to the continuity of tractions.
A convergence study is performed using the four different discretizations, fixing the element sizes to the values $h \in \{1/10, 1/20, 1/30, 1/40\}$. The considered discretizations are fitted to the spherical interface $\Gamma_I$, such that $\mathcal{T}_{h}(\Omega) = \mathcal{T}_{h}(\Omega_s) \sqcup \mathcal{T}_{h}(\Omega_f)$. The convergence results presented in Fig. \ref{fig:test2}b demonstrate a systematic reduction in the displacement and pressure errors under mesh refinement for both formulations, with rates comparable to those obtained in the single-domain verification cases. In particular, the Taylor–Hood $P_2/P_1$ formulation exhibits approximately third-order convergence in the displacement $L^2$--norm and second-order convergence in the pressure $L^2$--norm and displacement $H^1$--norm. The VMS-stabilized $P_1/P_1$ formulation achieves approximately second-order convergence in the displacement $L^2$--norm and first-order convergence in the displacement $H^1$-norm, while the pressure error decreases consistently with mesh refinement, approaching first-order convergence. These results verify the convergence behavior of the coupled formulation in the presence of the fluid--solid interface.

Figure \ref{fig:test2}c shows the comparison regarding the computational cost of the two formulations in terms of wall-clock time and pressure accuracy. For the same tetrahedral mesh, the $P_1/P_1$ discretization involves considerably fewer degrees of freedom than the $P_2/P_1$ formulation, since the displacement field is interpolated only at the mesh vertices. On the finest discretization, comprising approximately $290\times10^3$ tetrahedral elements, the wall-clock time is approximately $35$ sec for the stabilized formulation, compared with $400$ sec for the Taylor–Hood, corresponding to a speedup of approximately 11-fold. Although the higher-order formulation provides greater accuracy for a given mesh resolution, the pressure-error-versus-wall-clock-time comparison highlights the desired computational efficiency of the stabilized equal-order formulation, which achieves comparable pressure errors at substantially reduced computational cost. This trade-off between accuracy and computational cost motivates the use of the VMS-stabilized $P_1/P_1$ formulation for large-scale three-dimensional anatomical geometries.

\FloatBarrier
\subsection{Implementation on the brain domain}
\label{Sec:brain}

\begin{figure}[!htb]
\centering
\includegraphics[scale=.3]{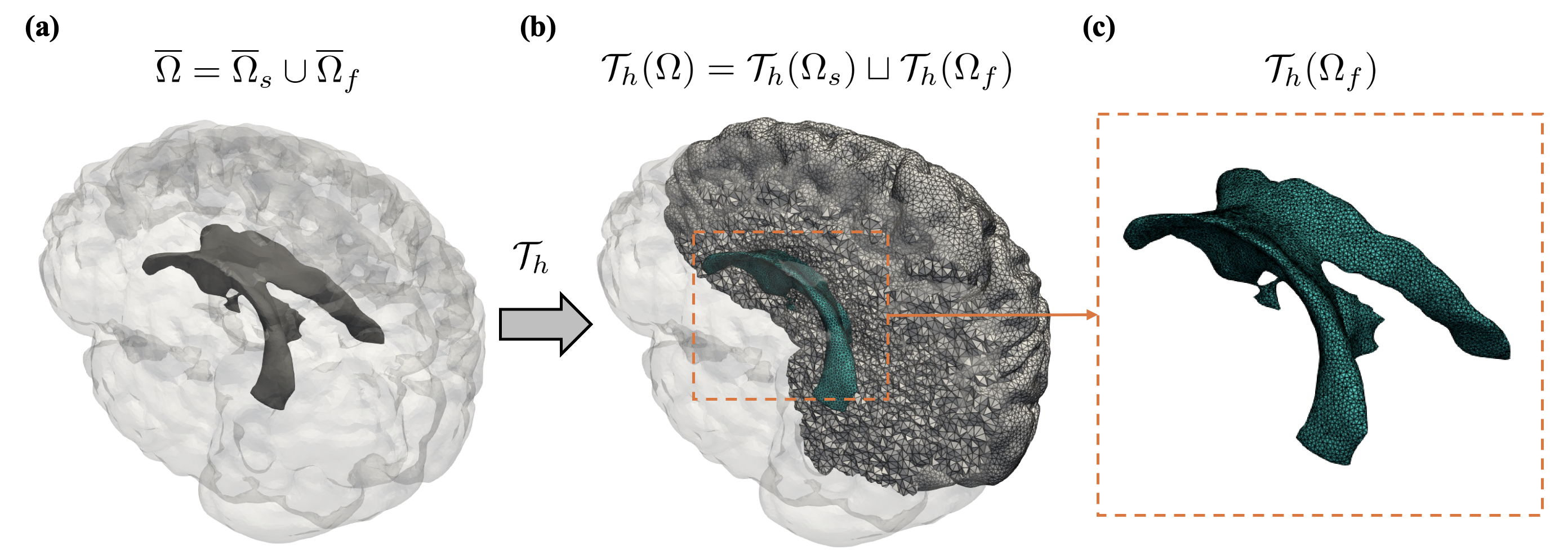} 
\caption{The anatomical brain--CSF configuration and computational discretization. (a) 3D representation of the brain parenchyma, $\Omega_s$, and the ventricular cerebrospinal fluid, $\Omega_f$. (b) The conforming tetrahedral finite element mesh of the brain–CSF domain, $\mathcal T_h(\Omega)=\mathcal T_h(\Omega_s)\sqcup\mathcal T_h(\Omega_f)$. (c) The detailed view of the tetrahedral mesh of the ventricular CSF domain, $\mathcal T_h(\Omega_f)$. }
\label{fig:model}
\end{figure}

In this stage, the proposed framework is implemented to simulate the wave propagation on an anatomical brain domain, obtained from an MRI-based geometrical reconstruction \cite{Tzirakis2023}.
Figure \ref{fig:model}a depicts the brain model which includes the brain parenchyma $\Omega_{s}$ as well as the lateral, third and fourth ventricles that represent $\Omega_f$.  The computational discretization $\mathcal{T}_{h}(\Omega) = \mathcal{T}_{h}(\Omega_s) \sqcup \mathcal{T}_{h}(\Omega_f)$ (Fig.~\ref{fig:model}b) consists of $784673$ tetrahedra ($746634$ in $\Omega_s$, $38039$ in $\Omega_f$), which are fitted to the brain-CSF surface $\Gamma_I$. The discretization parts $\mathcal{T}_{h}(\Omega_s)$, $\mathcal{T}_{h}(\Omega_f)$ are perfectly matched on $\Gamma_{I}$, i.e.: every facet of an element $E_{s} \in \mathcal{T}_{h}(\Omega_s)$, that is fitted on $\Gamma_{I}$, is shared with the corresponding facet of an element $E_{f} \in \mathcal{T}_{h}(\Omega_f)$ that is also fitted on $\Gamma_{I}$.

According to the configuration of Fig.~\ref{fig:bvp}, the excitation is applied as a harmonic displacement $\mathbf{g}=(u_0,0,0)$ with amplitude $u_0 = -5.5 \times 10^{-6} \text{m}$ on the two pad regions $\Gamma_{pad1}$ and $\Gamma_{pad2}$. The displacement is fixed to zero vector on $\Gamma_{neck}$, while the remaining outer boundary is traction-free. Numerical simulations are performed for the frequency values $f= \{ 15, 25, 35, 45 \}$~Hz. In the present study, the viscoelastic behavior of the brain parenchyma is described employing the Kelvin--Voigt model \cite{Fung1993}, with material parameters obtained by experimental MRE measurements reported in \cite{sack2008}. As shown in \cite{sack2008}, both the storage modulus $\mu_s'$ and the loss modulus $\mu_s''$ exhibit frequency-dependent behaviour. Since \cite{sack2008} reports the storage modulus $\mu_s'$ and viscosity $\eta_s$ at excitation frequencies of $f=25$ and $50$~Hz, linear fitting is employed to estimate these parameters for other frequencies. It should be noted that the employed viscoelastic parameters lead to a frequency-dependent material damping ratio $\tan\delta=\mu_s''/\mu_s'$, which estimates the relative contribution of viscous dissipation to elastic energy storage.
\begin{table}[ht]
  \centering
  \caption{Material parameters of the brain--CSF model.
  The frequency-dependent properties $\mu_s'$ and $\eta_s$ of the brain tissue are obtained by linear fitting on the results reported in \cite{sack2008}, with $f$ in Hz. Then, the loss modulus is calculated as $\mu_s'' = 2 \pi f \, \eta_{s}$.}
  \label{tab:params}
  \begin{tabular}{llll}
    \hline
    & Parameter & Value & Ref. \\
    \hline
    Brain parenchyma
      & storage modulus $\mu_s'(f)$
      & $(0.0156f+0.780)$ kPa
      & \cite{sack2008} \\
      & loss modulus $\mu_s''(f)$
      & $2\pi f(0.012f+2.8)\times 10^{-3}$ kPa
      & \cite{sack2008} \\
      & Poisson's ratio $\nu$
      & $0.45$ & \\
      & density $\rho_s$
      & $1000$ kg/m$^3$ & \\
    CSF
      & dynamic viscosity $\eta_f$
      & $0.001$ $\text{Pa}\cdot \text{sec}$  & \\
      & density $\rho_f$
      & $1000$ kg/m$^3$ & \\
    \hline
  \end{tabular}
\end{table}
For the considered excitation frequencies, the loss factor increases from approximately $0.28$ at $15$ Hz to $0.64$ at $45$ Hz, indicating an increasingly significant dissipative contribution at higher frequencies. The employed material parameters for both brain and CSF are listed in Table~\ref{tab:params}.

For comparison, a simplified model is also considered, in which the fluid subdomain is excluded and its mechanical effect on the surrounding brain tissue is approximated by a spatially uniform normal traction applied to the ventricular wall:
\begin{equation}
  \boldsymbol{\sigma}_s(\mathbf{u}_s,p_s) \cdot \mathbf{n} = -\bar{p}\,\mathbf{n}
  \qquad \text{on } \Gamma_I,
  \label{eq:const_traction}
\end{equation}
where $\bar{p}$ is a constant pressure. In order to prescribe a comparable pressure loading, the coupled brain--CSF problem is first solved for each excitation frequency $f$. Then, the pressure value $\bar{p}$ is determined as the maximum value of real part of the computed CSF pressure field $\Re(p_{f})$ and applied uniformly on $\Gamma_I$. This procedure provides a reference pressure load of comparable magnitude to that obtained from the coupled simulation.

The simplified model neglects two important effects captured by the coupled formulation. Firstly, the spatial variation of the CSF pressure on $\Gamma_I$ is disregarded. Secondly, the kinematic constraint associated with the incompressibility of the enclosed fluid is no longer enforced. In the absence of fluid sources or sinks and for an impermeable ventricular boundary, this constraint implies conservation of the ventricular volume, i.e.: $\int_{\Gamma_I}\mathbf{u}\cdot\mathbf{n}\,d\Gamma=0$. Consequently, the traction-based model allows deformations of the ventricular wall that are not subject to the global volume conservation constraint imposed by the enclosed incompressible CSF.

It should be emphasized that the uniform-traction model is intended as a simplified reference representation of the ventricular CSF, rather than an optimized approximation of the coupled fluid response. The prescribed pressure is selected from the maximum real pressure computed in the corresponding coupled simulation, and alternative choices could lead to different displacement fields. In particular, a suitably calibrated uniform pressure could potentially reproduce the global ventricular volume conservation condition. However, such a calibration would not generally recover the non-uniform and complex-valued traction distribution or the viscous and kinematic interactions associated with the explicitly resolved CSF. Therefore, the differences observed between the two models should be interpreted in relation to the particular uniform-traction approximation considered in this study, rather than as evidence that all possible constant-pressure approximations necessarily produce the same behaviour. 

\begin{figure}[!htb]
\centering
\includegraphics[scale=.34]{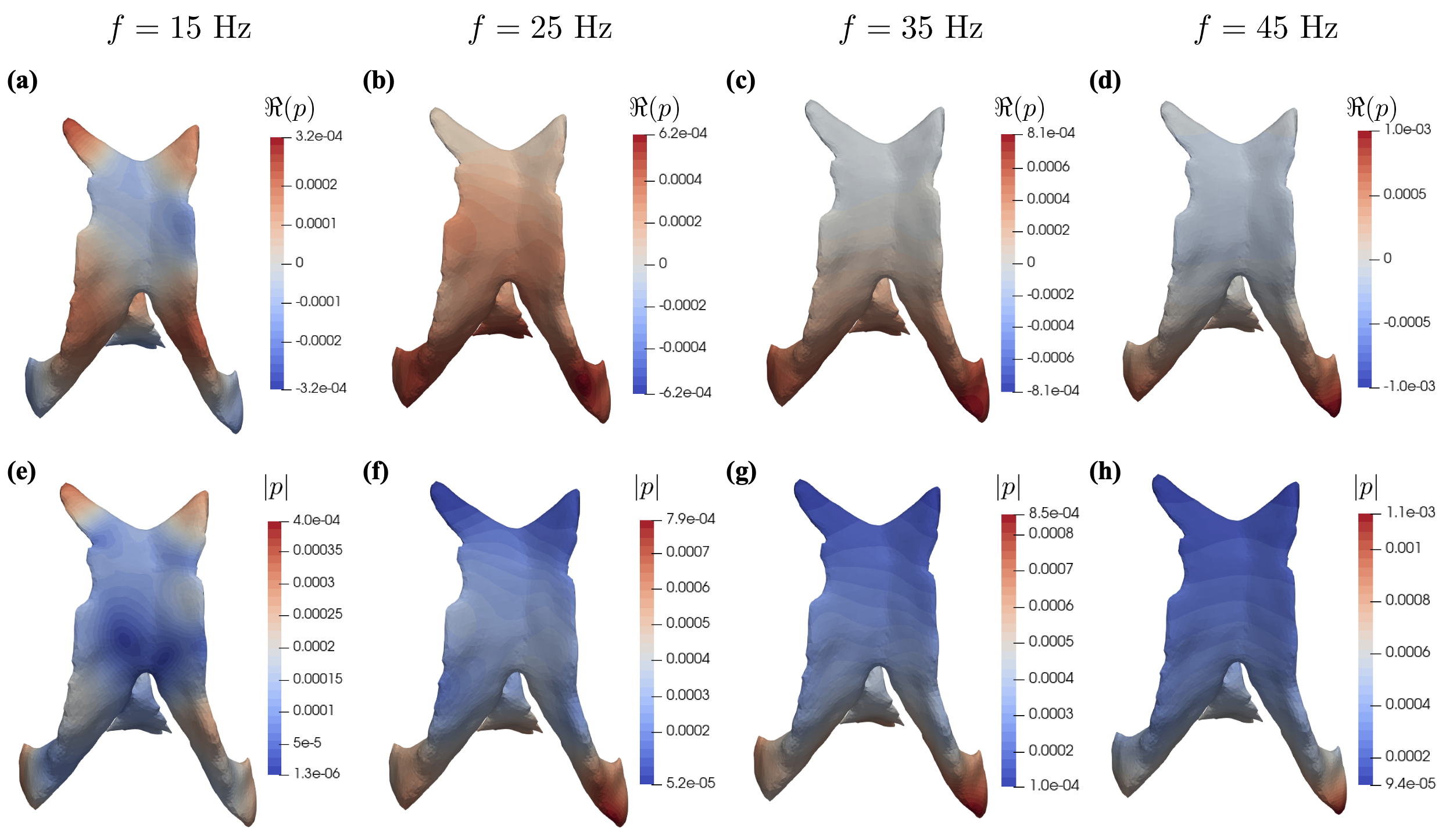} 
\caption{Frequency-dependent pressure distribution in the ventricular cerebrospinal fluid $\Omega_{f}$. Spatial distribution of (a-d) the real part $\Re(p_f)$ , and (e-h) the absolute value $|p_{f}| = \sqrt{p_f\overline{p}_f}$, obtained from the coupled brain--CSF model at the frequency range 15-45 Hz. The colour scales are adjusted independently to the range $[-\operatorname{max}\Re(p_{f}),\operatorname{max}\Re(p_{f})]$ for each frequency to highlight spatial variations in the pressure field, with symmetric limits around zero for the real part. }
\label{fig:csf_pressure}
\end{figure}

Figure \ref{fig:csf_pressure} presents the spatial distribution of the complex pressure field in the ventricular CSF for the excitation frequencies $f= \{ 15, 25, 35, 45 \}$~Hz. For the contour plots Fig. \ref{fig:csf_pressure}a-d, the colour scales are adjusted independently for each frequency in order to vary in the range $[-\operatorname{max}\Re(p_{f}),\operatorname{max}\Re(p_{f})]$. Both the real part $\Re(p_{f})$ and the absolute value $|p_{f}|=\sqrt{p_f\overline{p}_f}$ of the pressure exhibit non-uniform spatial distributions, which vary with excitation frequency. At 15 Hz, the real part of pressure reveals distributed pronounced regions of positive and negative signs across the ventricular system, whereas at higher frequencies, the distribution changes, with larger positive values concentrated toward the posterior ventricular regions. In addition, the pressure modulus reveals a redistribution of the response, with relatively larger values occurring near the posterior parts of the ventricular geometry at higher frequencies. Furthermore, the maximum values of $\Re(p_{f})$ and $|p_{f}|$ increase with increasing frequency. These results demonstrate the spatial variability of the pressure loading exerted by the CSF on the surrounding brain tissue, which is neglected when the fluid contribution is approximated by a spatially uniform normal traction.

\begin{figure}[!htb]
\centering
\includegraphics[scale=.46]{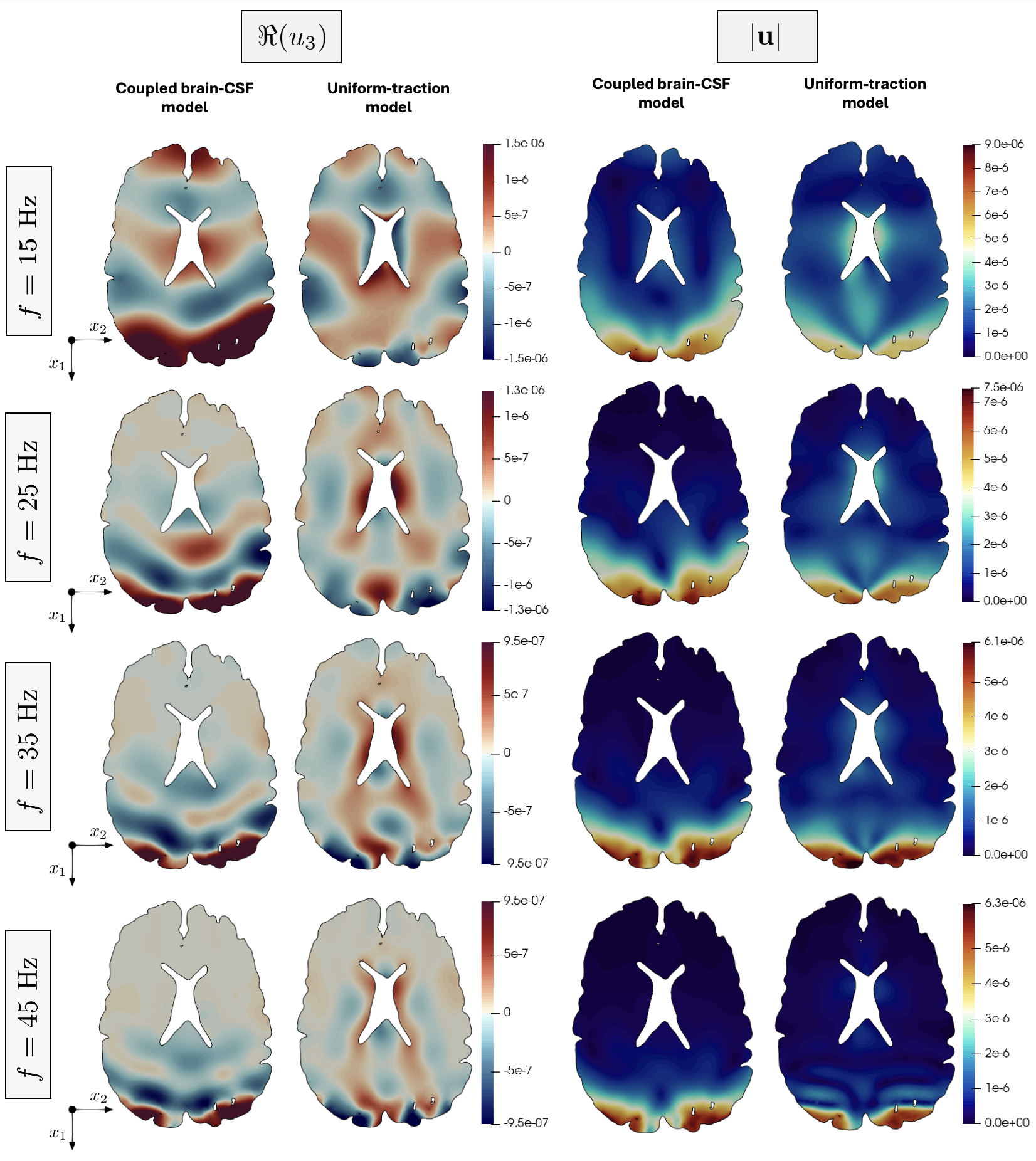} 
\caption{Comparison of the displacement fields obtained from the coupled brain–CSF model and the simplified model employing a uniform traction boundary condition on the ventricular walls. Spatial distributions of the magnitude of the real displacement component $\Re(u_3)$, as well as the norm of the complex displacement amplitude $|\mathbf{u}| = \sqrt{\mathbf{u} \cdot \overline{\mathbf{u}}}$, are shown in the horizontal plane at excitation frequencies $f= \{ 15, 25, 35, 45 \}$~Hz. In order to highlight spatial variations in the real part $\Re(u_3)$, the colour scales are adjusted to the range $[-\operatorname{max}\Re(u_3),\operatorname{max}\Re(u_3)]$ for each frequency, with $\operatorname{max}\Re(u_3)$ corresponding to the results of the uniform-traction model. For the $|\mathbf{u}|$, the colour scales are adjusted to the range $[0,\operatorname{max}|\mathbf{u}|]$, where $\operatorname{max}|\mathbf{u}|$ corresponds to the results the coupled model.}
\label{fig:umag_slices}
\end{figure}

Figure \ref{fig:umag_slices} compares the displacement fields obtained from the coupled brain–CSF formulation and the simplified model employing a uniform traction boundary condition on the ventricular walls. The real part of the displacement component $\Re(u_{3})$ and the absolute value of the complex displacement vector $|\mathbf{u}| = \sqrt{\mathbf{u} \cdot \overline{\mathbf{u}}}$ are presented on the horizontal plane $x_3=0.17$ for the brain parenchyma. For each excitation frequency, common colour scales are employed for the two models to facilitate their direct comparison.

\begin{figure}[!htb]
\centering
\includegraphics[scale=.48]{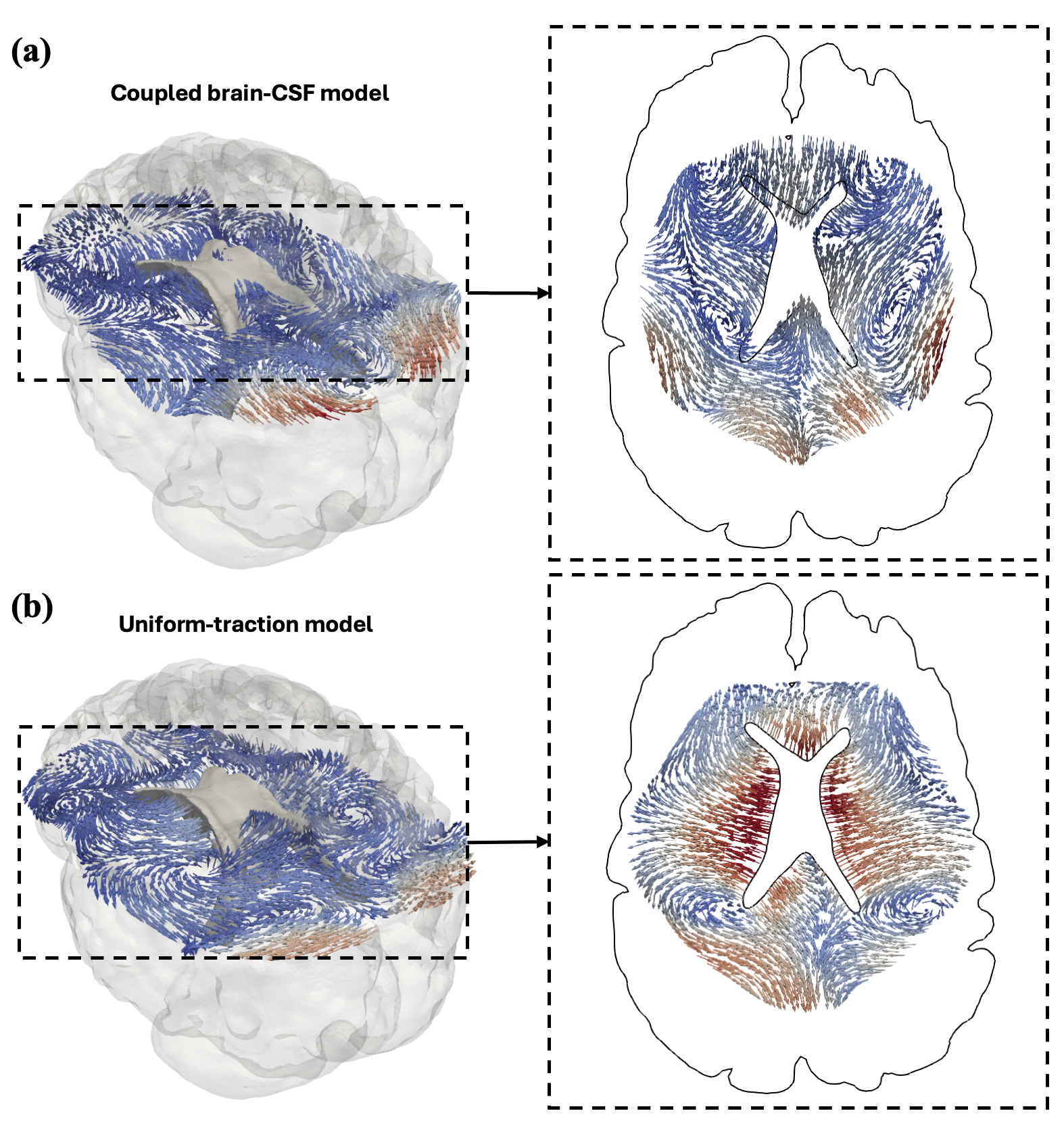} 
\caption{Comparison of the displacement vector fields obtained from: (a) the coupled brain–CSF model and (b) the simplified model employing a uniform traction boundary condition on the ventricular walls. The right column provides magnified views of the displacement fields in the vicinity of the ventricles, highlighting the different deformation patterns predicted by the two models.}
\label{fig:glyphs}
\end{figure}

For both modelling approaches, the displacement response becomes increasingly localized near the excitation boundary as the frequency increases, which is consistent with the attenuation associated with the adopted viscoelastic material properties. Nevertheless, differences in the spatial distribution of the displacement fields are observed between the two models, particularly in the regions surrounding the ventricles. The uniform-traction model exhibits more pronounced displacement amplitudes in the periventricular regions, particularly at the lower excitation frequencies. Furthermore, the spatial distribution of $\Re(u_3)$ reveals distinct patterns of positive and negative displacement between the two models across the considered frequency range. Interestingly, at 45 Hz, the distributions of the displacement modulus become qualitatively more similar, although differences in the real displacement component remain apparent. This observation indicates that a comparison based solely on the displacement modulus may not fully capture the influence of the brain–CSF interface treatment on the predicted wave field.

\begin{figure}[!htb]
\centering
\includegraphics[scale=.55]{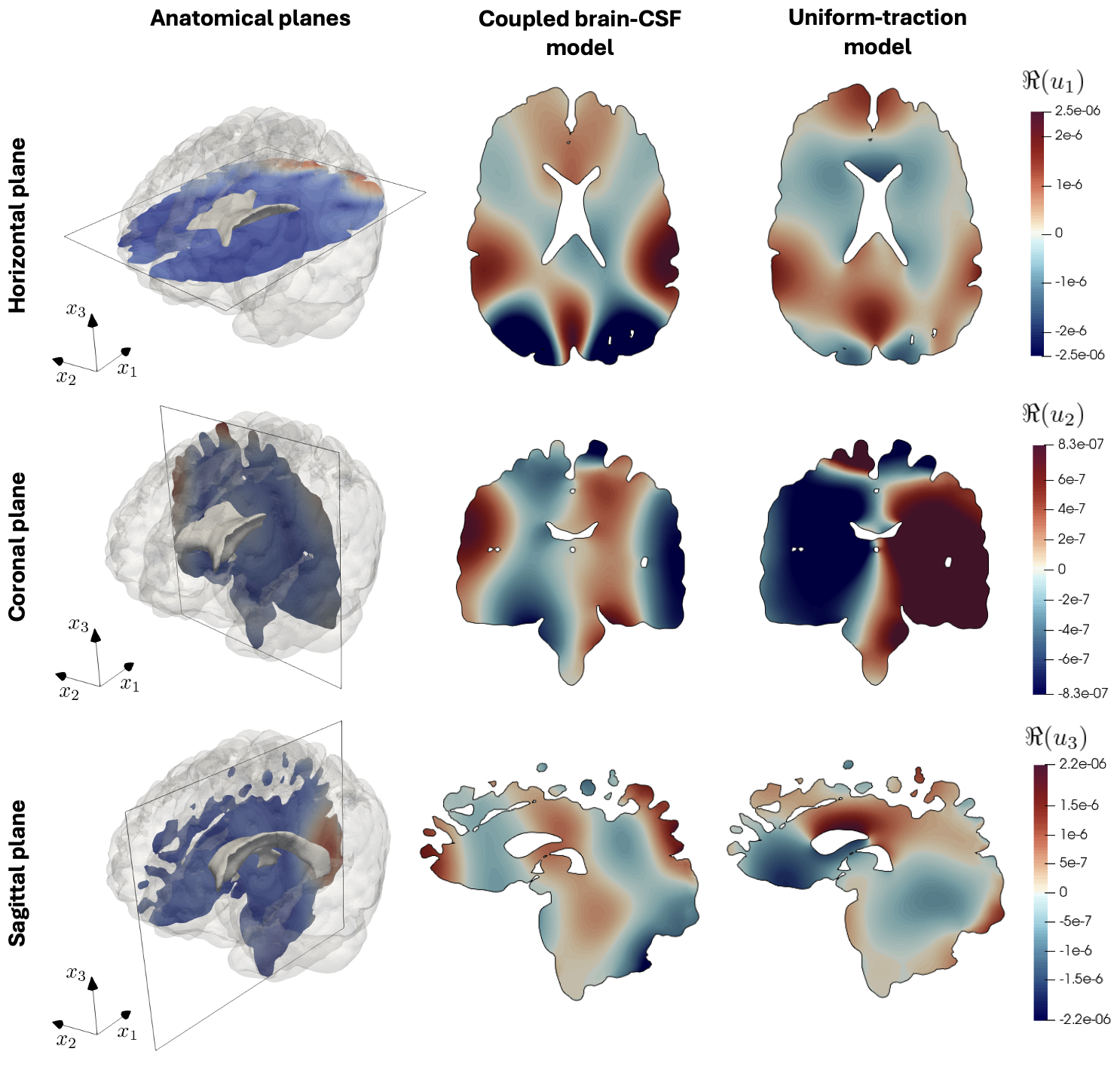} 
\caption{Comparison of the spatial distributions of the real part of displacement components obtained from the coupled brain--CSF model and the simplified model employing a uniform traction boundary condition on the ventricular walls, for the frequency $f = 15$ Hz. The real parts of the displacement components $\Re(u_1)$, $\Re(u_2)$ and $\Re(u_3)$ are presented in the horizontal, coronal, and sagittal planes, respectively. The left column illustrates the corresponding anatomical section planes, while the middle and right columns show the displacement fields predicted by the two models. In order to highlight spatial variations in the real part $\Re(u_3)$, the colour scales are adjusted to the range $[- \operatorname{max}\Re(u_j),\operatorname{max}\Re(u_j)]$, with $\operatorname{max}\Re(u_j)$ corresponding to the results of the coupled model. }
\label{fig:brain_all_planes}
\end{figure}

The different deformation patterns near the ventricular walls are further illustrated in Fig. \ref{fig:glyphs}, where the displacement vector fields obtained from both models are presented on the horizontal plane. In the coupled brain–CSF model (Fig. \ref{fig:glyphs}a), the displacement vectors tend to follow the ventricular boundary, whereas the uniform-traction model (Fig. \ref{fig:glyphs}b) exhibits more pronounced displacement components directed away from the ventricular walls. The latter deformation pattern is indicative of ventricular dilatation, consistent with the absence of an explicitly imposed ventricular volume conservation constraint in the simplified model.





In order to further investigate the influence of the CSF, Fig. \ref{fig:brain_all_planes} compares the real components of the displacement field for all three anatomical planes. The results show that different patterns of positive and negative displacement are formed. The displacement "dipoles" in the region that surrounds the ventricular boundary, especially for $\Re(u_{2})$ on the coronal plane and $\Re(u_{3})$ on the sagittal plane, explain the dilatation of the ventricles due to the uniform applied traction. The observed changes in the component-wise displacement fields further highlight the influence of the interface treatment on the spatial and phase-dependent mechanical response of the brain parenchyma.



It should be noted that the present investigation is based on several modelling assumptions. In particular, the brain parenchyma is represented as a homogeneous, isotropic, linear viscoelastic medium, while the CSF is modelled as an incompressible Newtonian fluid governed by the unsteady Stokes equations. Moreover, the simplified uniform traction model provides a reference for examining the influence of the brain--CSF interface treatment, rather than an exact approximation of the coupled fluid response. Consequently, the comparisons presented here provide a qualitative assessment of the effects of fluid--solid coupling on the predicted displacement fields. Further investigations incorporating spatially heterogeneous tissue properties as well as the presence of skull and CSF in the subarachnoid space would be valuable to assess the implications of these effects for the reconstruction of MRE-based parameters.

\FloatBarrier
\section{Conclusions}

In this work, a monolithic mixed $u-p$ finite element formulation has been developed and investigated for the modelling of the time-harmonic wave propagation in coupled viscoelastic solid--fluid domains, with particular application to brain magnetic resonance elastography. The proposed framework provides a uniform representation of the brain domain, considering viscoelastic behaviour for the brain parenchyma, while the cerebrospinal fluid is modelled as an incompressible fluid assuming Stokes flow. The boundary value problem is expressed in the frequency domain complemented by the corresponding kinematic and dynamic interface conditions. A residual-based stabilization derived from the Variational Multiscale Method (VMS) is introduced in the fluid subdomain to accommodate computationally efficient equal-order $P_1/P_1$ finite element interpolation functions.

The numerical framework has been verified through test cases, based on manufactured solutions, involving individual viscoelastic solid and Stokes flow problems, followed by a coupled configuration containing an embedded fluid inclusion. The results demonstrated systematic convergence of the displacement and pressure approximations, with the VMS-based $P_1/P_1$ formulation. Compared with the Taylor–Hood $P_2/P_1$ discretization, the stabilized equal-order formulation also demonstrated a substantial reduction in computational cost. In the finest coupled benchmark, comprising approximately $290000$ tetrahedral elements, the computational time decreased from approximately $400$ sec to $35$ sec,  highlighting the efficiency of the proposed approach for large-scale three-dimensional simulations.

The next step was the implementation of the proposed framework to an anatomical brain--CSF model, where the influence of ventricular CSF on the predicted wave propagation patterns has been revealed. The coupled simulations exhibited spatially non-uniform, frequency-dependent pressure distributions within the ventricular system. Comparisons with a simplified model employing a spatially uniform traction on the ventricular walls revealed differences in the displacement field, particularly in regions surrounding the ventricles. Furthermore, the simplified model exhibited deformation patterns associated with ventricular dilatation, whereas the coupled formulation accounts for the volume conservation constraint imposed by the enclosed incompressible fluid. These observations highlight the importance of the mechanical constraints introduced by the CSF in determining the displacement response of the surrounding brain tissue.

Overall, the proposed monolithic stabilized finite element framework provides a computationally efficient approach for investigating fluid--solid interactions, with application in brain MRE. The results indicate that replacing the ventricular CSF with a prescribed uniform traction may lead to different local displacement patterns, emphasizing the relevance of explicitly modelling the fluid compartment. While the present investigation considers a homogeneous linear viscoelastic representation of the brain tissue, the developed framework provides a basis for future work incorporating spatial variability of parenchyma properties as well as the presence of additional components to the model such as the skull and the subarachnoid space.

\section*{Appendix A} \label{appBVP}

\textbf{\large Time-to-frequency domain transformation and pressure discontinuity on brain/CSF interface}
\vspace{0.1cm}

\noindent
\underline{Time-to-frequency domain transformation of the boundary value problem}
\medskip

\noindent
Given the brain/CSF configuration $\Omega = \Omega_{s} \cup \Omega_{f} \cup \Gamma_{I}$, the set of the following time-domain partial differential equations is considered:
\begin{equation}
\begin{aligned}
\Div \Stress (\mathbf{u}_{s},p_{s}) - \rho_{s}\partial_{tt}\mathbf{u}_{s} &= \mathbf{0} \quad \text{in } \Omega_{s}
\\
\Div \Stress (\mathbf{v}_{f}, p_{f}) - \rho_{f}\partial_{t}\mathbf{v}_{f} &= \mathbf{0} \quad \text{in } \Omega_{f}
\\
\Div \mathbf{v}_{f} &= 0 \quad \text{in } \Omega_{f}
\end{aligned}
\label{Eq:TDpdes}
\end{equation}
along with the interface and boundary conditions:
\begin{equation}
\begin{aligned}
\partial_{t} \mathbf{u}_{s} = \mathbf{v}_{f}, \qquad \Stress (\mathbf{u}_{s},p_{s}) \cdot \nv_{s} + \Stress (\mathbf{v}_{f},p_{f}) \cdot \nv_{f} = \mathbf{0} \quad \text{on } \Gamma_{I}
\end{aligned}
\label{Eq:TDIntf}
\end{equation}
where $\mathbf{u}_{s}: \Omega_{s} \times [0,\infty) \rightarrow \mathbb{R}^{3}$ and $p_{s}: \Omega_{s} \times [0,\infty) \rightarrow \mathbb{R}$ are the time-domain displacement and pressure of brain, while $\mathbf{v}_{f}: \Omega_{f} \times [0,\infty) \rightarrow \mathbb{R}^{3}$ and $p_{f}: \Omega_{f} \times [0,\infty) \rightarrow \mathbb{R}$ denote the velocity and pressure of the CSF. Assuming that the brain tissue behaves as a linear and isotropic viscoelastic medium, following a Kelvin-Voigt model, the stress tensor is related with the strain and strain rate as:
\begin{equation}
\begin{aligned}
\Stress (\mathbf{u}_{s},p_{s}) &= -p_{s}\idem + 2\mu_{0} \mathbf{e}(\mathbf{u}_{s}) + 2\eta_{0} \, \mathbf{e}(\partial_{t} \mathbf{u}_{s})
\\
p_{s} &= -\lambda_{0} \Div \mathbf{u}_{s} - \zeta_{0} \Div (\partial_{t} \mathbf{u}_{s})
\end{aligned}
\label{Eq:StressKV}
\end{equation}
where $\eta_{0}, \zeta_{0}$ represent the viscous parameters of the Kelvin--Voigt model \cite{Fung1993}. As time-harmonic displacement field of the form $\mathbf{u}(\mathbf{x},t) = \Re \{ \hat{\mathbf{u}}(\mathbf{x},\omega) \, e^{i\omega t} \}$ is generated in brain tissue, the stress tensor reads:
\begin{equation}
\begin{aligned}
\Stress (\mathbf{u}_{s},p_{s}) &= \Re \Big\{ \big[ -\hat{p}_{s} \idem + 2(\mu_{0} + i \omega \eta_{0}) \,\mathbf{e}(\hat{\mathbf{u}}_{s}) \big] e^{i\omega t} \Big\}
\\
\hat{p}_{s} &= -(\lambda_{0}+i \omega \zeta_{0}) \Div \hat{\mathbf{u}}_{s}
\end{aligned}
\label{Eq:StressSFD}
\end{equation}
Similarly, based on the time-harmonic velocity $\mathbf{v}(\mathbf{x},t) = \Re \{ i \omega \, \hat{\mathbf{u}}_{f}(\mathbf{x},\omega) \, e^{i\omega t} \}$, and pressure $p_{f}(\mathbf{x},t) = \Re \{ \hat{p}_{f}(\mathbf{x},\omega) \, e^{i\omega t} \}$ fields in the CSF, the fluid stress tensor is written in terms of displacement $\hat{\mathbf{u}}_{f}$ and pressure $\hat{p}_{f}$ as:
\begin{equation}
\begin{aligned}
\Stress (\mathbf{u}_{f},p_{f}) &= \Re \Big\{ \big[ -\hat{p}_{f} \idem + 2 i \omega \eta_{f} \,\mathbf{e}(\hat{\mathbf{u}}_{f}) \big] e^{i\omega t} \Big\}
\end{aligned}
\label{Eq:StressFFD}
\end{equation}
Substituting the stress expressions \eqref{Eq:StressSFD} - \eqref{Eq:StressFFD} into the boundary value problem \eqref{Eq:TDpdes}-\eqref{Eq:TDIntf}, we obtain the frequency domain mixed boundary value problem \eqref{Eq:Unified} - \eqref{Eq:BCs},  provided in the main part of the text, where $\mu_{s} = \mu_s' + i \mu_s'' = \mu_{0} + i \omega \eta_{0}$ and $\lambda_{s} = \lambda' + i \lambda'' = \lambda_{0} + i \omega \zeta_{0}$. For brevity, the hat symbol is omitted for the complex amplitudes, i.e. $\mathbf{u}$, $p$ are used instead of $\hat{\mathbf{u}}$, $\hat{p}$.

\vspace{0.5cm}
\medskip
\noindent
\underline{Pressure discontinuity on brain/CSF interface}
\medskip

\noindent
As the pressure controls the normal part of the traction vector $\mathbf{t}_{n} = (\mathbf{t} \cdot \nv)\nv$, the dynamic interface condition \eqref{Eq:Interface} for the normal traction component takes the form:
\begin{equation}
\begin{aligned}
p_{s} - p_{f} = 2\kappa_{s} \, e_{nn}(\mathbf{u}_{s}) - 2\kappa_{f} \, e_{nn}(\mathbf{u}_{f})
\end{aligned}
\label{Eq:NormalInterface}
\end{equation}
where $e_{nn}(\mathbf{u}) = \nv \cdot \Strain \cdot \nv = \nv \cdot \Grad \mathbf{u} \cdot \nv$. Based on the definition of the surface divergence $\Div_{\Gamma} \mathbf{u} = \big(\idem - \nv \otimes \nv \big):\Grad \mathbf{u} = \Div \mathbf{u} - e_{nn}(\mathbf{u})$, the continuity of the displacement \eqref{Eq:Interface}, implies continuity for the $\Div_{\Gamma}$ as well. Therefore, the continuity of the surface divergence gives:
\begin{equation}
\begin{aligned}
e_{nn}(\mathbf{u}_{s}) = \Div \mathbf{u}_{s} +  e_{nn}(\mathbf{u}_{f})
\end{aligned}
\label{Eq:SurfDiv}
\end{equation}
Substituting Eq. \eqref{Eq:SurfDiv} into Eq. \eqref{Eq:NormalInterface} and considering that $p_{s} = - \lambda_{s} \Div \mathbf{u}_{s}$, the normal component of the traction interface condition leads to the following pressure relation:
\begin{equation}
\begin{aligned}
\Big ( 1 + 2\frac{\kappa_{s}}{\lambda_{s}} \Big) p_{s} - p_{f} = 2 (\kappa_{s} - \kappa_{f}) \, e_{nn}(\mathbf{u})
\end{aligned}
\label{Eq:Pdisc}
\end{equation}
The only reason for pressure continuity is $\frac{\kappa_{s}}{\lambda_{s}} p_{f} =(\kappa_{s} - \kappa_{f}) \, e_{nn}(\mathbf{u})$, as stated in Section \ref{Sec:StrongForm}. It should be noted that in the incompressible elastic case for the brain ($\lambda_{s} \rightarrow \infty$), the continuity is guaranteed when even if $\mu_{s} = i \omega \eta_{f}$ or $e_{nn}=0$.

\vspace{1cm}

\section*{Appendix B} \label{appVMS}

\textbf{\large Variational Multiscale Method}
\vspace{0.1cm}

\noindent
The mixed displacement–pressure formulation considered in this work involves an incompressibility constraint in the fluid subdomain, for which standard Galerkin discretizations employing equal-order interpolation functions, such as $P_1/P_1$, may exhibit spurious pressure modes. To address this issue, the present work adopts the Variational Multiscale (VMS) framework introduced by Hughes \cite{Hughes1995}, following the approach developed by Masud and co-workers for problems in fluid and solid mechanics \cite{Masud2009, masud2005}. The VMS approach accounts for the influence of unresolved solution scales on the finite element approximation through their local representation and elimination. In the present formulation, this procedure is applied to the time-harmonic Stokes equations in the fluid subdomain, employing complex-valued coefficients and element-level bubble functions to approximate the unresolved displacement scales. 

For the derivation of the VMS stabilization, we consider
a generic time-harmonic incompressible problem defined
on a domain $\Omega$, with a prescribed traction
$\mathbf{t}_N$ on the Neumann boundary $\partial\Omega_N$.
The corresponding mixed weak formulation reads:
\begin{equation}
\begin{aligned}
a(\mathbf{w},\mathbf{u})-b(\mathbf{w},p)
&= \int_{\partial\Omega_N}
\mathbf{w}\cdot\mathbf{t}_N\,d\Gamma,\\
b(\mathbf{u},\nu)&=0,
\end{aligned}
\label{eq:vms_generic_weak}
\end{equation}
where the $a(\cdot,\cdot)$ and $b(\cdot,\cdot)$ denote the following bilinear forms:
\begin{equation}
\begin{aligned}
a(\mathbf{w},\mathbf{u}) = \int_{\Omega} 2 \kappa \, \mathbf{e}(\mathbf{w}):\mathbf{e}(\mathbf{u})\,  d\Omega - \int_{\Omega} \rho \omega^{2} \, \mathbf{w} \cdot \mathbf{u}\,  d\Omega , \qquad
b(\mathbf{w},\upsilon) = \int_{\Omega}(\Div \mathbf{w})\upsilon \, d\Omega
\end{aligned}
\label{Eq:Bilin}
\end{equation}
The derivation of the stabilized weak formulation begins with the decomposition of the displacement and the corresponding vector test function into a coarse (resolved) and a fine (unresolved) part:
\begin{equation}
\mathbf{u} = \mathbf{u}_{h} + \mathbf{u}', \qquad \mathbf{w} = \mathbf{w}_{h} + \mathbf{w}'
\end{equation}
while the pressure is represented only on the coarse scale, $p = p_{h}$, $\upsilon = \upsilon_{h}$. The fine scales $\mathbf{u}'$, $\mathbf{w}'$ are defined piecewise-continuous functions that vanish on element boundaries. Substituting the scale decomposition into
Eq.~\eqref{eq:vms_generic_weak}, the weak formulation splits into a coarse-scale and a fine-scale sub-problem, i.e:

\medskip
\noindent \textit{Coarse-scale sub-problem:}
\begin{equation}
\begin{aligned}
a(\mathbf{w}_{h},\mathbf{u}_{h}) - b(\mathbf{w}_{h},p_{h}) &= - a(\mathbf{w}_{h},\mathbf{u}') + \int_{\partial \Omega_{N}} \mathbf{w}_{h} \cdot \mathbf{t}_{N} \, d \Gamma, \\
b(\mathbf{u}_{h},\upsilon_{h}) + b(\mathbf{u}',\upsilon_{h}) &= 0,
\end{aligned}
\label{Eq:Coarse}
\end{equation}

\medskip
\noindent \textit{Fine-scale sub-problem:}
\begin{equation}
\begin{aligned}
a(\mathbf{w}',\mathbf{u}') = -a(\mathbf{w}',\mathbf{u}_{h}) + b(\mathbf{w}',p_{h})
\end{aligned}
\label{Eq:Fine}
\end{equation}
The fine scales are represented locally in each element $E$ by a bubble function $b_{E}$,
\begin{equation}
\mathbf{u}'|_{E} = b_{E}(\boldsymbol{\xi})\,\hat{\mathbf{m}}, \qquad \mathbf{w}'|_{E} = b_{E}(\boldsymbol{\xi})\,\hat{\mathbf{l}}, \qquad \hat{\mathbf{m}},\hat{\mathbf{l}} \in \mathbb{C}^{3}
\label{Eq:Bubble}
\end{equation}
where, for linear tetrahedra, $b_{E} = 256\,\lambda_{1}\lambda_{2}\lambda_{3}\lambda_{4}$ in terms of the barycentric coordinates $\lambda_{i}$. Since $\mathbf{w}'$ is zero on $\partial E$, integrating by parts the right-hand side of~\eqref{Eq:Fine} over $E$ gives:
\begin{equation}
\begin{aligned}
a(\mathbf{w}', \mathbf{u}_{h}) - b(\mathbf{w}', p_{h}) = -\big(\mathbf{w}',\mathcal{R}(\mathbf{u}_{h}, p_{h}) \big)
\end{aligned}
\label{Eq:FineRhs}
\end{equation}
where $\mathcal{R}$ denotes the residual of the coarse-scale solution in the momentum equation:
\begin{equation}
\begin{aligned}
\big(\mathbf{w}',\mathcal{R}(\mathbf{u}_{h}, p_{h}) \big) = \int_{\Omega} \mathbf{w}' \cdot \mathcal{R}(\mathbf{u}_{h}, p_{h}) \, d\Omega, \qquad \mathcal{R}(\mathbf{u}_{h}, p_{h}) = 2\kappa \, \Div \mathbf{e}(\mathbf{u}_{h}) - \nabla p_{h} + \rho\omega^{2}\mathbf{u}_{h}
\end{aligned}
\label{Eq:CoarseResid}
\end{equation}
Then the fine-scale sub-problem takes the form:
\begin{equation}
\begin{aligned}
a(\mathbf{w}', \mathbf{u}') = \big(\mathbf{w}',\mathcal{R}(\mathbf{u}_{h}, p_{h}) \big)
\end{aligned}
\label{Eq:Fine2}
\end{equation}
Considering the fine-scale definitions \eqref{Eq:Bubble} and assuming that $\mathcal{R}(\mathbf{u}_{h}, p_{h})$ is constant inside $E$, Eq. \eqref{Eq:Fine2} is solved:
\begin{equation}
\mathbf{u}' = \boldsymbol{\tau} \cdot \mathcal{R}(\mathbf{u}_{h},p_{h}),
\qquad
\boldsymbol{\tau}(\mathbf{x}) = b_{E}(\mathbf{x}) \left( \int_{E} b_{E}\, d\Omega \right) \mathbf{K}_{E}^{-1},
\label{Eq:FineSol}
\end{equation}
with
\begin{equation}
\mathbf{K}_{E} = \int_{E} \kappa \left( |\nabla b_{E}|^{2}\,\mathbf{I} + \nabla b_{E}\otimes\nabla b_{E} \right) d\Omega - \rho\omega^{2}\int_{E} b_{E}^{2}\, d\Omega\; \mathbf{I}.
\label{Eq:Tau}
\end{equation}
The stabilization tensor $\boldsymbol{\tau}$ is complex-valued, symmetric, and computed element by element from the material parameters and the element geometry; no user-defined stabilization parameter is involved. 
As $\mathbf{u}'=\mathbf{0}$ on $\partial E$, the integration by parts of the right-hand terms of~\eqref{Eq:FineRhs} gives:
\begin{equation}
\begin{aligned}
&a(\mathbf{w}_{h},\mathbf{u}') = -\sum_{E \in \mathcal{T}_{h}} \int_{E} \big( 2\kappa \, \Div \mathbf{e}(\mathbf{w}_{h}) + \rho \omega^{2}\mathbf{w}_{h} \big) \cdot \mathbf{u}' d \Omega
\\
&b(\mathbf{u}',\upsilon_{h}) =  -\sum_{E \in \mathcal{T}_{h}} \int_{E} \nabla \upsilon_{h} \cdot \mathbf{u}' d \Omega
\end{aligned}    
\label{Eq:CoarseRhs}
\end{equation}
Substituting Eq. \eqref{Eq:FineSol} into the coarse-scale sub-problem \eqref{Eq:Coarse} along with updated right-hand side terms \eqref{Eq:CoarseRhs}, the stabilized weak formulation takes the form:

\medskip
\noindent Find $(\mathbf{u}_{h},p_{h}) \in \mathcal{U}_{h}\times\mathcal{Q}_{h}$ such that
\begin{equation}
\begin{aligned}
&\int_{\Omega} 2\kappa\, \mathbf{e}(\mathbf{w}_{h}) : \mathbf{e}(\mathbf{u}_{h})\,d\Omega -
\int_{\Omega} \rho \, \omega^2 \, \mathbf{w}_{h} \cdot \mathbf{u}_{h}\,d\Omega -
\int_{\Omega} (\Div \mathbf{w}_{h})\, p_{h}\,d\Omega - 
\\
&\int_{\Omega} (\Div \mathbf{u}_{h})\, \upsilon_{h}\,d\Omega - 
\sum_{E\in\mathcal{T}_{h}^{f}} 
\int_{E}
\boldsymbol{\tau}\,\mathcal{R}_{f}(\mathbf{w}_{h},\upsilon_{h}) \cdot \mathcal{R}_{f}(\mathbf{u}_{h},p_{h}) \, d\Omega = \int_{\partial \Omega} \mathbf{w}\cdot\mathbf{h}\, d\Gamma    
\end{aligned}
\label{Eq:VMS2}
\end{equation}
for all $(\mathbf{w}_{h},\upsilon_{h}) \in \mathcal{U}_{0,h}\times\mathcal{Q}_{h}$.
\medskip

\noindent
The stabilization term is symmetric in its two arguments and vanishes for the exact solution, so the method is consistent. Employing linear interpolation functions (P1) for the displacement which leads to $\Div \mathbf{e}(\mathbf{u}_{h}) = \mathbf{0}$ and assuming piecewise constant $\kappa$ in each element, the residual reduces to $\mathcal{R}(\mathbf{u}_{h},p_{h}) = \rho\omega^{2}\mathbf{u}_{h} - \nabla p_{h}$. In matrix form, the stabilized system becomes
\begin{equation}
\begin{bmatrix} \mathbf{A} + \mathbf{S}_{uu} & \mathbf{B}^{T}\! - \mathbf{S}_{up}^{T} \\ \mathbf{B} - \mathbf{S}_{up} & -\mathbf{S}_{pp} \end{bmatrix}
\cdot
\begin{Bmatrix} \mathbf{u}_{h} \\ \mathbf{p}_{h} \end{Bmatrix}
=
\begin{Bmatrix} \mathbf{f} \\ \mathbf{0} \end{Bmatrix},
\label{Eq:VMSMatrix}
\end{equation}
with element contributions
\begin{equation}
\mathbf{S}_{uu}^{E} = -\int_{E} \rho^{2}\omega^{4}\, \mathbf{N}^{T}\boldsymbol{\tau}\,\mathbf{N}\, d\Omega, \qquad
\mathbf{S}_{up}^{E} = -\int_{E} \rho\omega^{2}\, (\nabla N)^{T}\boldsymbol{\tau}\,\mathbf{N}\, d\Omega, \qquad
\mathbf{S}_{pp}^{E} = \int_{E} (\nabla N)^{T}\boldsymbol{\tau}\,\nabla N\, d\Omega,
\label{Eq:StabBlocks}
\end{equation}
where $\mathbf{N}$ and $N$ denote the displacement and pressure shape functions. 

\vspace{1cm}

\section*{Acknowledgements}
D.C.R.\ acknowledges the support from the project \emph{BioMagnus} (OPPTY-MSCA/0125/0002) which is implemented through the "HORIZON EUROPE- 2nd  Opportunity" Programme, under the programme of social cohesion “THALIA 2021-2027”, co-funded by the European Union through Research and Innovation Foundation (RIF) of Cyprus. 
I.S.\ acknowledges the financial support from the German Research Foundation (project: 460333672 CRC 1540 EBM).
V.V.\ acknowledges the support from EISMEA for the EIC Pathfinder Open project \emph{IgnitePLASMA} (grant ID: 101129853).






\bibliographystyle{unsrt}
\bibliography{refs}

@article{muthupillai1995,
  author  = {Muthupillai, R. and Lomas, D. J. and Rossman, P. J. and Greenleaf, J. F. and Manduca, A. and Ehman, R. L.},
  title   = {Magnetic resonance elastography by direct visualization of propagating acoustic strain waves},
  journal = {Science},
  volume  = {269},
  number  = {5232},
  pages   = {1854--1857},
  year    = {1995},
  doi     = {10.1126/science.7569924}
}

@book{hirsch2017book,
  author    = {Hirsch, Sebastian and Braun, J{\"u}rgen and Sack, Ingolf},
      title     = {Magnetic Resonance Elastography: Physical Background and Medical Applications},
  publisher = {Wiley-VCH},
  address   = {Weinheim},
  year      = {2017}
}

@article{manduca2021,
  author  = {Manduca, Armando and Bayly, Philip V. and Ehman, Richard L. and Kolipaka, Arunark and Royston, Thomas J. and Sack, Ingolf and Sinkus, Ralph and Van Beers, Bernard E.},
  title   = {{MR} elastography: Principles, guidelines, and terminology},
  journal = {Magnetic Resonance in Medicine},
  volume  = {85},
  number  = {5},
  pages   = {2377--2390},
  year    = {2021}
}

@article{hiscox2016,
  author  = {Hiscox, Lucy V. and Johnson, Curtis L. and Barnhill, Eric and McGarry, Matt D. J. and Huston, John and van Beek, Edwin J. R. and Starr, John M. and Roberts, Neil},
  title   = {Magnetic resonance elastography ({MRE}) of the human brain: technique, findings and clinical applications},
  journal = {Physics in Medicine and Biology},
  volume  = {61},
  number  = {24},
  pages   = {R401--R437},
  year    = {2016}
}

@article{McGarry2015,
  author  = {McGarry, Matthew D. J. and Johnson, Curtis L. and Sutton, Bradley P. and Georgiadis, John G. and Van Houten, Elijah E. W. and Pattison, Adam J. and Weaver, John B. and Paulsen, Keith D.},
  title   = {Suitability of poroelastic and viscoelastic mechanical models for high and low frequency {MR} elastography},
  journal = {Medical Physics},
  volume  = {42},
  number  = {2},
  pages   = {947--957},
  year    = {2015},
  doi     = {10.1118/1.4905048}
}

@article{murphy2019,
  author  = {Murphy, Matthew C. and Huston, John and Ehman, Richard L.},
  title   = {{MR} elastography of the brain and its application in neurological diseases},
  journal = {NeuroImage},
  volume  = {187},
  pages   = {176--183},
  year    = {2019}
}

@article{sack2009,
  author  = {Sack, Ingolf and Beierbach, Bernd and Wuerfel, Jens and Klatt, Dieter and Hamhaber, Uwe and Papazoglou, Sebastian and Martus, Peter and Braun, J{\"u}rgen},
  title   = {The impact of aging and gender on brain viscoelasticity},
  journal = {NeuroImage},
  volume  = {46},
  number  = {3},
  pages   = {652--657},
  year    = {2009}
}

@article{sack2008,
author = {Sack, Ingolf and Beierbach, Bernd and Hamhaber, Uwe and Klatt, Dieter and Braun, Jürgen},
title = {Non-invasive measurement of brain viscoelasticity using magnetic resonance elastography},
journal = {NMR in Biomedicine},
volume = {21},
number = {3},
pages = {265-271},
doi = {https://doi.org/10.1002/nbm.1189},
url = {https://analyticalsciencejournals.onlinelibrary.wiley.com/doi/abs/10.1002/nbm.1189},
year = {2008}
}

@article{wuerfel2010,
  author  = {Wuerfel, Jens and Paul, Friedemann and Beierbach, Bernd and Hamhaber, Uwe and Klatt, Dieter and Papazoglou, Sebastian and Zipp, Frauke and Martus, Peter and Braun, J{\"u}rgen and Sack, Ingolf},
  title   = {{MR}-elastography reveals degradation of tissue integrity in multiple sclerosis},
  journal = {NeuroImage},
  volume  = {49},
  number  = {3},
  pages   = {2520--2525},
  year    = {2010}
}

@article{murphy2011,
  author  = {Murphy, Matthew C. and Huston, John and Jack, Clifford R. and Glaser, Kevin J. and Manduca, Armando and Felmlee, Joel P. and Ehman, Richard L.},
  title   = {Decreased brain stiffness in {A}lzheimer's disease determined by magnetic resonance elastography},
  journal = {Journal of Magnetic Resonance Imaging},
  volume  = {34},
  number  = {3},
  pages   = {494--498},
  year    = {2011}
}

@article{streitberger2011,
  author  = {Streitberger, Kaspar-Josche and Wiener, Edzard and Hoffmann, Jan and Freimann, Florian Baptist and Klatt, Dieter and Braun, J{\"u}rgen and Lin, Kui and McLaughlin, Joyce and Sprung, Christian and Klingebiel, Randolf and Sack, Ingolf},
  title   = {In vivo viscoelastic properties of the brain in normal pressure hydrocephalus},
  journal = {NMR in Biomedicine},
  volume  = {24},
  number  = {4},
  pages   = {385--392},
  year    = {2011}
}

@article{lilaj2021,
  author  = {Lilaj, Ledia and Herthum, Helge and Meyer, Tom and Shahryari, Mehrgan and Bertalan, Gergely and Caiazzo, Alfonso and Braun, J{\"u}rgen and Fischer, Thomas and Hirsch, Sebastian and Sack, Ingolf},
  title   = {Inversion-recovery {MR} elastography of the human brain for improved stiffness quantification near fluid--solid boundaries},
  journal = {Magnetic Resonance in Medicine},
  volume  = {86},
  number  = {5},
  pages   = {2552--2561},
  year    = {2021}
}

@article{lilaj2020,
  author  = {Lilaj, Ledia and Fischer, Thomas and Guo, Jing and Braun, J{\"u}rgen and Sack, Ingolf and Hirsch, Sebastian},
  title   = {Separation of fluid and solid shear wave fields and quantification of coupling density by magnetic resonance poroelastography},
  journal = {Magnetic Resonance in Medicine},
  volume  = {85},
  number  = {3},
  pages   = {1655--1668},
  year    = {2021}
}

@article{perrinez2009,
  author  = {Perri{\~n}ez, Phillip R. and Kennedy, Francis E. and Van Houten, Elijah E. W. and Weaver, John B. and Paulsen, Keith D.},
  title   = {Modeling of soft poroelastic tissue in time-harmonic {MR} elastography},
  journal = {IEEE Transactions on Biomedical Engineering},
  volume  = {56},
  number  = {3},
  pages   = {598--608},
  year    = {2009}
}

@article{galarce2023,
  author  = {Galarce, Felipe and Tabelow, Karsten and Polzehl, J{\"o}rg and Papanikas, Christos Panagiotis and Vavourakis, Vasileios and Lilaj, Ledia and Sack, Ingolf and Caiazzo, Alfonso},
  title   = {Displacement and pressure reconstruction from magnetic resonance elastography images: Application to an in silico brain model},
  journal = {SIAM Journal on Imaging Sciences},
  volume  = {16},
  number  = {2},
  pages   = {996--1027},
  year    = {2023}
}

@article{mcgrath2016,
  author  = {McGrath, Deirdre M. and Ravikumar, Nishant and Wilkinson, Iain D. and Frangi, Alejandro F. and Taylor, Zeike A.},
  title   = {Magnetic resonance elastography of the brain: An in silico study to determine the influence of cranial anatomy},
  journal = {Magnetic Resonance in Medicine},
  volume  = {76},
  number  = {2},
  pages   = {645--662},
  year    = {2016},
  doi     = {10.1002/mrm.25881}
}

@book{Fung1993,
  author    = {Fung, Yuan-Cheng},
  title     = {Biomechanics: Mechanical Properties of Living Tissues},
  edition   = {2},
  year      = {1993},
  publisher = {Springer},
  address   = {New York},
  doi       = {10.1007/978-1-4757-2257-4},
  isbn      = {978-1-4757-2257-4}
}

@article{Tzirakis2023,
  author  = {Tzirakis, Konstantinos and
             Papanikas, Christos Panagiotis and
             Sakkalis, Vangelis and
             Tzamali, Eleftheria and
             Papaharilaou, Yannis and
             Caiazzo, Alfonso and
             Stylianopoulos, Triantafyllos and
             Vavourakis, Vasileios},
  title   = {An adaptive semi-implicit finite element
             solver for brain cancer progression modeling},
  journal = {International Journal for Numerical Methods
             in Biomedical Engineering},
  volume  = {39},
  number  = {7},
  pages   = {e3734},
  year    = {2023},
  doi     = {10.1002/cnm.3734}
}

@article{Fovargue2018Review,
  author  = {Fovargue, Daniel and Nordsletten, David and Sinkus, Ralph},
  title   = {Stiffness reconstruction methods for {MR} elastography},
  journal = {NMR in Biomedicine},
  volume  = {31},
  number  = {10},
  pages   = {e3935},
  year    = {2018},
  doi     = {10.1002/nbm.3935}
}

@article{Fovargue2018Reconstruction,
  author  = {Fovargue, Daniel and Kozerke, Sebastian and Sinkus, Ralph and Nordsletten, David},
  title   = {Robust {MR} elastography stiffness quantification using a localized divergence free finite element reconstruction},
  journal = {Medical Image Analysis},
  volume  = {44},
  pages   = {126--142},
  year    = {2018},
  doi     = {10.1016/j.media.2017.12.005}
}

@book{Boffi2013,
  author    = {Boffi, Daniele and Brezzi, Franco and Fortin, Michel},
  title     = {Mixed Finite Element Methods and Applications},
  publisher = {Springer},
  address   = {Berlin, Heidelberg},
  year      = {2013}
}

@article{taylorhood1973,
  author  = {Taylor, C. and Hood, P.},
  title   = {A numerical solution of the {N}avier--{S}tokes equations using the finite element technique},
  journal = {Computers \& Fluids},
  volume  = {1},
  number  = {1},
  pages   = {73--100},
  year    = {1973}
}

@article{hughes1986,
  author  = {Hughes, Thomas J. R. and Franca, Leopoldo P. and Balestra, Marc},
  title   = {A new finite element formulation for computational fluid dynamics: {V}. {C}ircumventing the {B}abu{\v{s}}ka--{B}rezzi condition: a stable {P}etrov--{G}alerkin formulation of the {S}tokes problem accommodating equal-order interpolations},
  journal = {Computer Methods in Applied Mechanics and Engineering},
  volume  = {59},
  number  = {1},
  pages   = {85--99},
  year    = {1986}
}

@article{codina2000,
  author  = {Codina, Ramon},
  title   = {Stabilization of incompressibility and convection through orthogonal sub-scales in finite element methods},
  journal = {Computer Methods in Applied Mechanics and Engineering},
  volume  = {190},
  number  = {13--14},
  pages   = {1579--1599},
  year    = {2000}
}

@article{Hughes1995,
  author  = {Hughes, Thomas J. R.},
  title   = {Multiscale phenomena: {G}reen's functions, the {D}irichlet-to-{N}eumann formulation, subgrid scale models, bubbles and the origins of stabilized methods},
  journal = {Computer Methods in Applied Mechanics and Engineering},
  volume  = {127},
  pages   = {387--401},
  year    = {1995},
  doi     = {10.1016/0045-7825(95)00844-9}
}

@Article{Masud2009,
author={Masud, Arif
and Calderer, Ramon},
title={A variational multiscale stabilized formulation for the incompressible Navier--Stokes equations},
journal={Computational Mechanics},
year={2009},
month={Jul},
day={01},
volume={44},
number={2},
pages={145-160},
doi={10.1007/s00466-008-0362-3},
url={https://doi.org/10.1007/s00466-008-0362-3}
}

@article{hughes1998,
  author  = {Hughes, Thomas J. R. and Feij{\'o}o, Gonzalo R. and Mazzei, Luca and Quincy, Jean-Baptiste},
  title   = {The variational multiscale method---a paradigm for computational mechanics},
  journal = {Computer Methods in Applied Mechanics and Engineering},
  volume  = {166},
  number  = {1--2},
  pages   = {3--24},
  year    = {1998}
}

@article{masud2005,
  author  = {Masud, Arif and Xia, Kaiming},
  title   = {A stabilized mixed finite element method for nearly incompressible elasticity},
  journal = {Journal of Applied Mechanics},
  volume  = {72},
  number  = {5},
  pages   = {711--720},
  year    = {2005}
}

@article{mcgrath2017delivery,
  author  = {McGrath, Deirdre M. and Ravikumar, Nishant and Beltrachini, Leandro and Wilkinson, Iain D. and Frangi, Alejandro F. and Taylor, Zeike A.},
  title   = {Evaluation of wave delivery methodology for brain {MRE}: Insights from computational simulations},
  journal = {Magnetic Resonance in Medicine},
  volume  = {78},
  number  = {1},
  pages   = {341--356},
  year    = {2017},
  doi     = {10.1002/mrm.26333}
}

@article{jiang2011,
  author  = {Jiang, Yu and Fujiwara, Hiroshi and Nakamura, Gen},
  title   = {Approximate Steady State Models for Magnetic Resonance Elastography},
  journal = {SIAM Journal on Applied Mathematics},
  volume  = {71},
  number  = {6},
  pages   = {1965--1989},
  year    = {2011},
  doi     = {10.1137/100781882}
}

@article{codina2002,
  author  = {Codina, Ramon},
  title   = {Stabilized finite element approximation of transient incompressible flows using orthogonal subscales},
  journal = {Computer Methods in Applied Mechanics and Engineering},
  volume  = {191},
  number  = {39--40},
  pages   = {4295--4321},
  year    = {2002},
  doi     = {10.1016/S0045-7825(02)00337-7}
}

@article{codinaBaiges2011,
  author  = {Codina, Ramon and Baiges, Joan},
  title   = {Finite element approximation of transmission conditions in fluids and solids introducing boundary subgrid scales},
  journal = {International Journal for Numerical Methods in Engineering},
  volume  = {87},
  number  = {1--5},
  pages   = {386--411},
  year    = {2011},
  doi     = {10.1002/nme.3111}
}

@article{rodopoulos2025vpinn,
  author  = {Rodopoulos, Dimitrios C. and Pantidis, Panos
             and Karathanasopoulos, Nikolaos},
  title   = {Variational {PINNs} with tree-based integration
             and boundary element data in the modeling of
             multi-phase architected materials},
  journal = {Engineering Analysis with Boundary Elements},
  volume  = {180},
  pages   = {106448},
  year    = {2025},
  doi     = {10.1016/j.enganabound.2025.106448}
}

@article{Carcamo2024,
  title   = {A stabilized total pressure-formulation of the
             Biot's poroelasticity equations in frequency domain:
             Numerical analysis and applications},
  author  = {Cárcamo, Cristian and Caiazzo, Alfonso
             and Galarce, Felipe and Mura, Joaquín},
  journal = {Computer Methods in Applied Mechanics and Engineering},
  volume  = {432},
  pages   = {117353},
  year    = {2024},
  doi     = {10.1016/j.cma.2024.117353}
}

@article{Galarce2021,
  title   = {Fast reconstruction of 3D blood flows from
             Doppler ultrasound images and reduced models},
  author  = {Galarce, Felipe and Gerbeau, Jean-Frédéric
             and Lombardi, Damiano and Mula, Olga},
  journal = {Computer Methods in Applied Mechanics and Engineering},
  volume  = {375},
  pages   = {113559},
  year    = {2021},
  doi     = {10.1016/j.cma.2020.113559}
}

@misc{Guennebaud2010,
  author       = {Guennebaud, Ga{\"e}l and
                  Jacob, Beno{\^i}t and others},
  title        = {{Eigen}},
  url ={\url{https://libeigen.gitlab.io}},
  year         = {2010}
}

@techreport{Balay2025,
  author      = {Balay, Satish and Abhyankar, Shrirang
                 and Adams, Mark F. and others},
  title       = {{PETSc/TAO} Users Manual},
  institution = {Argonne National Laboratory},
  number      = {ANL-21/39 - Revision 3.24},
  year        = {2025},
  doi         = {10.2172/2998643}
}

@article{dagum1998openmp,
  author  = {Dagum, Leonardo and Menon, Ramesh},
  title   = {{OpenMP}: An industry standard {API} for
             shared-memory programming},
  journal = {IEEE Computational Science and Engineering},
  volume  = {5},
  number  = {1},
  pages   = {46--55},
  year    = {1998},
  doi     = {10.1109/99.660313}
}

@book{saad2003iterative,
  author    = {Saad, Yousef},
  title     = {Iterative Methods for Sparse Linear Systems},
  edition   = {2nd},
  publisher = {Society for Industrial and Applied Mathematics},
  address   = {Philadelphia},
  year      = {2003},
  doi       = {10.1137/1.9780898718003}
}

\end{document}